\documentclass[twocolumn]{aastex631}
\usepackage{captcont}
\usepackage{amsmath}
\usepackage{multirow}
\usepackage{mathtools}

\submitjournal{ApJ}

\shorttitle{VLITE HUDF}
\shortauthors{Polisensky et al.}

\begin{document}

\title{VLITE Observations of the {\it Hubble} Ultra Deep Field}

\correspondingauthor{Emil Polisensky}
\email{emil.j.polisensky.civ@us.navy.mil}

\author[0000-0003-3272-9237]{Emil Polisensky}
\affiliation{U.S.\ Naval Research Laboratory, Code 7213, 4555 Overlook Ave SW, Washington, DC 20375, USA}

\author[0000-0002-5187-7107]{Wendy Peters}
\affiliation{U.S.\ Naval Research Laboratory, Code 7213, 4555 Overlook Ave SW, Washington, DC 20375, USA}

\author[0000-0002-1634-9886]{Simona Giacintucci}
\affiliation{U.S.\ Naval Research Laboratory, Code 7213, 4555 Overlook Ave SW, Washington, DC 20375, USA}

\author[0000-0003-1991-370X]{Kristina Nyland}
\affiliation{U.S. Naval Observatory, 3450 Massachusetts Avenue NW, Washington, DC 20392, USA}

\author[0000-0001-6812-7938]{Tracy Clarke}
\affiliation{U.S.\ Naval Research Laboratory, Code 7213, 4555 Overlook Ave SW, Washington, DC 20375, USA}

\author[0009-0008-0527-4082]{Amaris V. McCarver}
\affiliation{Department of Astronomy, University of Virginia, 530 McCormick Road, Charlottesville, VA 22904, USA}
\affiliation{National Radio Astronomy Observatory, 520 Edgemont Road, Charlottesville, VA 22903, USA}

\begin{abstract}

We present 136~hr of 340~MHz observations of the {\it Hubble} Ultra Deep Field (HUDF) conducted with the VLA Low-band Ionosphere and Transient Experiment (VLITE), collected commensally with higher-frequency Karl G. Jansky Very Large Array (VLA) observations. Our deepest image combines 20~hr of data, covers 4~deg$^2$, and achieves an angular resolution of $8.8^{\prime\prime} \times 3.8^{\prime\prime}$ with an rms sensitivity of 158~$\mu$Jy~bm$^{-1}$ at the field center. We catalog 671 radio sources detected at $\geq 5\sigma$ significance. Correcting for incompleteness using injection-recovery simulations with a model radio continuum population, we show that our source counts recover the sub-mJy star-forming galaxy population at abundances consistent with other deep fields at comparable frequencies. The data span 45 observing days across two VLA A-configuration semesters separated by 1.3~years, providing a search for variability on $\sim$1-year timescales at sub-mJy flux densities. We identify eleven variable sources with flux densities of 1--100~mJy and modulation indices of 6--40\%, six of which exhibit peaked radio spectra consistent with compact AGN. Their variability amplitudes are consistent with refractive interstellar scintillation (RISS) predicted for this line of sight, though three sources with modulation indices exceeding 30\% likely require an intrinsic contribution. One source, subsequently confirmed as a compact symmetric object through VLBA follow-up observations, demonstrates the potential of VLITE for identifying populations of young radio AGN through commensal low-frequency variability searches.

\end{abstract}

%% See the online documentation for the f\newcommand*{\logten}{\mathop{\log_{10}}}ull list of available subject keywords and the rules for their use.

\section{Introduction} \label{sec:intro}

Deep radio continuum surveys are essential for studying the complex interplay between supermassive black hole (SMBH) growth and stellar assembly across cosmic time. Observations at low radio frequencies ($<1$~GHz) are particularly sensitive to older populations of cosmic ray electrons associated with high-redshift radio active galactic nuclei (AGN), which exhibit steep radio spectral indices \citep{tielens+79, miley2008}, as well as to diffuse, extended emission from radio galaxy lobes that fades rapidly at higher frequencies. Although low-frequency surveys play a pivotal role in understanding the evolution of radio source populations, the calibration and imaging of deep, wide-field, high-angular-resolution sub-GHz data remain technically and computationally challenging. Nevertheless, recent advances have enabled increasingly sensitive surveys around 325 MHz (P-band), with several achieving rms noise levels below 1~mJy~bm$^{-1}$ (see Table~\ref{tab:Pband_surveys}).

A novel approach to deep low-frequency radio surveys leverages commensal systems that utilize the infrastructure of radio interferometers operating at higher frequencies. The VLA Low-band Ionosphere and Transient Experiment \citep[VLITE;][]{clarke+16, polisensky+16} exemplifies such a system, operating on the Karl G. Jansky Very Large Array (VLA). VLITE continuously records a 64~MHz bandwidth spanning 320--384~MHz from the VLA low-band system using a dedicated DiFX software correlator and automated pipelines for data processing, cataloging, and archiving. VLITE science operations began on 25 November 2014 with 10 antennas and have since expanded to 18, operating commensally with nearly all VLA observations\footnote{Exceptions include NRAO TAC-approved P-band projects and observations of moving targets (e.g., solar system objects) at any band.} and accumulating over 6200~hr of data annually. All VLITE data products are hosted by the U.S. Naval Research Laboratory (NRL), representing a largely untapped resource for enhancing the legacy value and scientific impact of VLA surveys.

The {\it Hubble} Ultra Deep Field \citep[HUDF;][]{beckwith+06} is an ideal target for deep low-frequency imaging. As one of the most extensively studied regions of the sky (see Section~\ref{sec:HUDF}), the HUDF benefits from an exceptional wealth of ancillary data spanning X-ray through infrared wavelengths, making it an ideal field for cross-identification and physical characterization of radio sources across a wide range of redshifts and source types. The multi-epoch nature of the VLITE observations, spanning two VLA A-configuration semesters separated by 1.3~years, additionally enables a search for low-frequency variability on year-long timescales, a largely unexplored regime sensitive to both refractive interstellar scintillation (RISS) of compact sources \citep{rickett86, rickett90, walker98} and intrinsic variability in AGN jets \citep{nyland+20, ross+21}.

In this paper, we present new high-resolution 340~MHz imaging of the HUDF using data from the VLITE commensal observing system. Our deepest image achieves a 1$\sigma$ rms sensitivity of 158~$\mu$Jy~bm$^{-1}$ over 4~deg$^2$, from which we catalog 671 radio sources. We show that the corrected source counts recover the sub-mJy star-forming galaxy (SFG) population at abundances consistent with other deep P-band fields, and we identify eleven variable sources whose properties are consistent with compact AGN undergoing RISS or intrinsic jet variability, including one source subsequently confirmed as a compact symmetric object through VLBA follow-up \citep{nyland+26}. Section~\ref{sec:HUDF} provides an overview of the HUDF field, followed by a description of the observations, data reduction, and imaging procedures in Section~\ref{sec:obs}. The resulting radio source catalog is presented in Section~\ref{sec:catalog}, and its properties, including completeness, spectral indices, and variability, are analyzed in Section~\ref{sec:analysis}. We conclude with a summary in Section~\ref{sec:summary}. Throughout this work, we define the spectral index $\alpha$ according to $S \propto \nu^\alpha$, where $S$ is flux density and $\nu$ is frequency.

\section{The Hubble Ultra-Deep Field}
\label{sec:HUDF}

The HUDF and its surrounding vicinity, including the GOODS-South\footnote{Great Observatories Origins Deep Survey-South \citep{dickinson+03}.} (GOODS-S) field, constitute one of the most extensively studied areas of the sky. Optical and near-infrared datasets from over 800 {\it Hubble} Space Telescope (HST) orbits \citep{beckwith+06} provide precise constraints on stellar population properties such as mass and age. Complementary observations from {\it Spitzer} \citep{labbe+15} and {\it Herschel} \citep{elbaz+11} cover the entire GOODS-S footprint, yielding measurements of dust content and star formation histories across a wide range of redshifts. Deep infrared imaging has been further enhanced by {\it James Webb} Space Telescope (JWST) observations of the HUDF using the NIRCam and NIRSpec instruments \citep{eisenstein+26}, which probe ultra-faint galaxies to the highest known redshifts ($z \gtrsim 13$) and can potentially identify hosts of rare sources with AGN signatures that remain undetected even in deep {\it Chandra} X-ray or VLA radio imaging. At higher energies, a total of 7~Ms (81~days) of combined {\it Chandra} \citep{xue+11} and {\it XMM-Newton} \citep{ranalli+13} observations provide deep X-ray imaging to identify unobscured, luminous AGN and constrain key SMBH properties such as bolometric luminosity.

The HUDF has also been observed extensively at radio frequencies. At 1.4~GHz, the Australia Telescope Large Area Survey \citep[ATLAS;][]{atlas1400} covers the GOODS-S region with a resolution of $16.3^{\prime\prime} \times 6.8^{\prime\prime}$ and a sensitivity of 14~$\mu$Jy~bm$^{-1}$, providing the highest-fidelity existing radio catalog of the field. At higher frequencies, deep VLA S-band (2--4~GHz) imaging of the HUDF \citep{rujopakarn+16, rujopakarn+18, alberts+20, adams+26} has reached sub-$\mu$Jy sensitivity, revealing the radio continuum emission from star-forming galaxies and AGN out to high redshift. The commensal VLITE observations presented here complement these higher-frequency datasets by providing simultaneous 340~MHz coverage, extending the spectral baseline to low frequencies where steep-spectrum AGN and diffuse radio emission are brightest.

\begin{table*}[t]
\caption{Summary of deep surveys at P-band. References: 1) \citealt{riseley+16}, 2) \citealt{sirothia+09}, 3) \citealt{owen+09}, 4) \citealt{sirothia+10}, 5) \citealt{smolcic+14}, 6) this work }
\label{tab:Pband_surveys}
{\footnotesize
\begin{tabular}{ccccccccc}
\hline
\hline
Field & Telescope & $\nu_{\rm central}$ & $\Delta \nu$ & $\sigma_{\rm rms}$ & $\theta_{\rm FWHM}$ & Area & $N$ & Ref. \\
 & & (MHz) & (MHz) & ($\mu$Jy~beam$^{-1}$) & (arcsec) & (deg$^2$) & & \\
(1) & (2) & (3) & (4) & (5) & (6) & (7) & (8) & (9)\\
\hline
Super-CLASS & GMRT & 325.0 & 33.0 & 34  & 13.00 $\times$ 13.00 & 6.50 & 3257 & 1 \\
ELAIS-N1 & GMRT & 325.0 & 32.0 & 40  & 9.35 $\times$ 7.38 & 3.80 & 1286 & 2 \\ 
SWIRE & VLA  & 324.5 & 11.7 & 70  & 6.37 $\times$ 5.90 & 3.14 & 1436 & 3 \\
ECDFS & GMRT & 325.0 & 32.0 & 75  & 10.80 $\times$ 7.30 & 1.54 & 667  & 4 \\
COSMOS & VLA  & 324.1 & 6.0  & 500 & 8.00 $\times$ 6.00 & 3.14 & 182  & 5 \\
HUDF & VLITE & 340.9 & 38.2 & 158 & 8.81 $\times$ 3.81 & 4.00 & 671 & 6 \\
\hline \hline
\end{tabular}}
\end{table*}

\section{Observations}
\label{sec:obs}

The VLA observed the HUDF at 2--4~GHz in its most extended (A) configuration for a total of 69~hr between April and June 2018 (project 18A-199) and an additional 100~hr between August and September 2019 (project 19A-242). The target field was centered at $\alpha_{\rm J2000} = 03^{\rm h} 32^{\rm m} 38.601^{\rm s}$, $\delta_{\rm J2000} = -27^{\circ}46^{\prime}58.83^{\prime\prime}$. Observing sessions consisted of 3--4~hr blocks that included scans of 3C48 for primary flux calibration and a nearby source for phase referencing. VLITE operated on up to 16 antennas during all but two observing blocks. The final set of observing dates and integration times for which VLITE data are available is summarized in Table~\ref{tab:18A-199}.

\subsection{VLITE Pipeline Data Reduction}
\label{sec:reduction}

VLITE data are transferred daily to computing systems at NRL, where they are processed in bulk using an automated pipeline that integrates Python with Obit \citep{cotton+08} and AIPS \citep{greisen+03} tasks. Because the frequency and spatial response of the 340~MHz receivers varies with the VLA primary observing band, data are grouped by primary observing frequency prior to processing. Approximately 24~hr of data are processed at a time. Scheduling blocks spanning 0h UTC are not split; instead, the entire dataset is processed on the start date, resulting in one image per UTC observation day even when two scheduling blocks were observed. Primary calibration may include data from other programs that observed any of VLITE's designated calibrators (3C138, 3C147, 3C196, 3C286, 3C295, 3C380) in addition to 3C48 observed by the HUDF program. VLITE does not utilize phase reference calibrators; scans intended for this purpose by the primary observers are treated as ``target'' fields in the subsequent analysis.

The processing pipeline begins by flagging known static radio frequency interference (RFI) signals, reducing the effective bandwidth from 64~MHz to 33.6~MHz centered at 338.5~MHz. The bulk of this frequency loss is due to persistent interference from satellite downlinks above 360~MHz. Remaining data are filtered in frequency and time using the Obit tasks \texttt{AutoFlag} and \texttt{MednFlag} to remove narrow-band and short-duration RFI. Calibrator scans with elevated rms noise are flagged in frequency. The AIPS task \texttt{RFLAG} is then applied to both cross-hand and parallel-hand data to remove broader spectral and temporal features, including signals that bleed into adjacent channels from satellite transmissions.

The calibrator data are used to solve for delay, complex bandpass, and gains using models distributed with Obit and referenced to the \citet{perley+17} flux scale. Based on prior experience with VLITE observations, the daily pipeline images are expected to reproduce this flux scale to an accuracy of $\sim$15\% (1$\sigma$), and we verify this expectation for the HUDF dataset in Section~\ref{sec:calib}. Solutions are filtered to remove outliers and smoothed to one value per calibrator scan. With these solutions applied, additional flagging clips calibrator values above 300~Jy and removes visibilities flagged more than 10\%. The data are then re-filtered to eliminate residual short-duration and narrow-band interference, and the calibration solutions are recomputed. A final editing step again removes visibilities flagged more than 10\%, after which the primary calibration solutions and all flags are applied to the data.

After primary calibration, all target fields are phase calibrated to sky models derived from the NVSS \citep{condon+98}, and calibrator fields are phase calibrated to their own models. Solutions are filtered and smoothed, with interpolation over any failed solutions, providing phase corrections accurate enough to proceed to imaging.

\subsection{Pipeline Imaging}
\label{sec:imaging}

Each target and calibrator field was first imaged without CLEANing to assess image statistics, from which final imaging parameters were derived based on integration time and array configuration. Data were imaged and self-calibrated using the Obit task \texttt{MFImage}, which divides the data into three frequency sub-bands for simultaneous deconvolution and applies direction-independent calibration. For the HUDF dataset, images covered a $1^\circ$ radius with additional small facet images placed on outlier sources within $3^\circ$.

CLEANing proceeded iteratively: an initial CLEAN to a 0.1~Jy residual threshold was followed by phase self-calibration with 12-second solution intervals. A second cycle applied deeper CLEANing with the same solution interval, culminating in a final CLEAN with 6000 components. The final image was accepted when the rms improved relative to the dirty image.

Table~\ref{tab:18A-199} summarizes the VLITE pipeline image properties for each day of projects 18A-199 and 19A-242. The achieved rms sensitivity is typically within a factor of $\sim$2 of the thermal noise\footnote{For reference, the VLA Exposure Calculator yields a thermal noise of 488~$\mu$Jy~bm$^{-1}$ for a 1-hour observation with 16 VLITE antennas and 30~MHz bandwidth.}, with $1.6 < \sigma_{\rm image}/\sigma_{\rm thermal} < 2.9$ across all epochs. The pipeline images achieve peak dynamic ranges of a few thousand and are not limited by dynamic range.

\startlongtable
\begin{deluxetable*}{ccccccccc}
\tablecaption{Summary of VLITE Observations \label{tab:18A-199}}
\tablecolumns{8}
\tablehead{
\colhead{Date} & \colhead{Time} & \colhead{$\sigma_{\rm thermal}$} & \colhead{$\sigma_{\rm image}$} & \colhead{$S_{\rm max}$} & \colhead{DR}  & \colhead{$\theta_{\rm maj} \times \theta_{\rm min}$} & \colhead{BPA} \\
\colhead{} & \colhead{(hr)} & \colhead{($\mu$Jy~bm$^{-1}$)} & \colhead{($\mu$Jy~bm$^{-1}$)} & \colhead{(Jy~bm$^{-1}$)} & \colhead{} & \colhead{($^{\prime \prime}$ $\times$ $^{\prime \prime}$)} & \colhead{($^o$)} \\
\colhead{(1)} & \colhead{(2)} & \colhead{(3)} & \colhead{(4)} & \colhead{(5)} & \colhead{(6)} & \colhead{(7)} & \colhead{(8)}
}
\startdata
\multicolumn{8}{c}{Daily pipeline images} 
\\
\hline
2018-04-29 &   1.59 &        387 &        752 &    1.51 &  2014 &  8.64 $\times$ 3.62 &  -9.82 \\ 
2018-04-30 &   2.22 &        328 &        691 &    1.49 &  2152 &  8.55 $\times$ 3.58 &  10.09 \\ 
2018-05-04 &   2.39 &        316 &        785 &    1.53 &  1952 &  8.78 $\times$ 3.58 &  13.55 \\ 
2018-05-05 &   1.59 &        387 &        823 &    1.51 &  1836 &  9.51 $\times$ 3.49 &  18.53 \\ 
2018-05-06 &   2.71 &        296 &        514 &    1.49 &  2896 &  8.29 $\times$ 3.60 &   5.48 \\ 
2018-05-07 &   2.46 &        311 &        513 &    1.53 &  2987 &  8.43 $\times$ 3.57 &   5.72 \\ 
2018-05-08 &   2.06 &        340 &        614 &    1.53 &  2489 &  8.73 $\times$ 3.51 &  12.11 \\ 
2018-05-12* &  3.47 &        262 &        421 &    1.53 &  3637 &  8.24 $\times$ 3.62 &   6.37 \\ 
2018-05-13 &   1.12 &        461 &        923 &    1.48 &  1602 &  8.54 $\times$ 3.52 &  -7.16 \\ 
2018-05-14* &  3.47 &        262 &        439 &    1.46 &  3322 &  8.17 $\times$ 3.63 &   3.04 \\ 
2018-05-17 &   2.47 &        311 &        522 &    1.45 &  2775 &  8.34 $\times$ 3.56 &   4.14 \\ 
2018-05-18 &   1.59 &        387 &        636 &    1.57 &  2468 &  9.46 $\times$ 3.49 &  18.22 \\ 
2018-05-19 &   2.47 &        311 &        718 &    1.44 &  2007 &  8.59 $\times$ 3.60 &   7.20 \\ 
2018-05-20* &  3.46 &        262 &        431 &    1.44 &  3338 &  8.18 $\times$ 3.62 &   4.08 \\ 
2018-05-25 &   2.46 &        311 &        631 &    1.55 &  2452 &  8.30 $\times$ 3.42 &   2.09 \\ 
2018-05-27 &   3.57 &        258 &        602 &    1.53 &  2542 & 10.00 $\times$ 3.58 &   7.40 \\ 
2018-05-28 &   3.17 &        274 &        802 &    1.47 &  1832 &  8.12 $\times$ 3.50 &   1.80 \\ 
2018-06-02 &   3.47 &        262 &        575 &    1.50 &  2614 &  8.14 $\times$ 3.47 &   3.30 \\ 
2018-06-03 &   3.46 &        262 &        526 &    1.33 &  2520 &  6.77 $\times$ 3.96 &  23.68 \\ 
2018-06-04 &   2.47 &        311 &        686 &    1.42 &  2071 &  8.85 $\times$ 3.57 &  12.46 \\ 
2018-06-10 &   2.47 &        311 &        794 &    1.50 &  1886 &  8.37 $\times$ 3.43 &   8.66 \\ 
2019-08-06 &   3.43 &        263 &        455 &    1.58 &  3470 &  8.11 $\times$ 4.49 &   5.72 \\ 
2019-08-17 &   3.47 &        262 &        655 &    0.59 &   908 &  7.52 $\times$ 4.44 &   2.78 \\ 
2019-08-18 &   3.44 &        263 &        680 &    0.61 &   898 &  7.87 $\times$ 4.34 &  -0.35 \\ 
2019-08-20 &   3.45 &        263 &        680 &    0.58 &   852 &  7.44 $\times$ 4.00 &   2.38 \\ 
2019-08-22 &   3.44 &        263 &        663 &    0.61 &   924 &  7.84 $\times$ 4.28 &  -2.55 \\ 
2019-08-23 &   3.44 &        263 &        726 &    0.61 &   841 &  7.68 $\times$ 3.98 &  -4.54 \\ 
2019-08-24 &   3.45 &        263 &        625 &    0.59 &   950 &  7.58 $\times$ 4.30 &  -0.43 \\ 
2019-08-25 &   3.45 &        263 &        588 &    0.63 &  1076 &  7.59 $\times$ 4.24 &   0.06 \\ 
2019-08-26 &   3.43 &        263 &        578 &    0.63 &  1086 &  7.62 $\times$ 4.32 &   3.13 \\ 
2019-08-27 &   3.45 &        263 &        603 &    0.62 &  1021 &  8.12 $\times$ 4.32 &   1.22 \\ 
2019-08-29 &   3.45 &        263 &        442 &    1.67 &  3776 &  8.99 $\times$ 4.86 &  10.87 \\ 
2019-08-30 &   3.43 &        263 &        419 &    1.63 &  3891 &  8.21 $\times$ 4.48 &   2.25 \\ 
2019-08-31 &   3.43 &        263 &        718 &    0.62 &   866 &  8.01 $\times$ 4.50 &   9.16 \\ 
2019-09-01 &   3.45 &        263 &        790 &    0.65 &   825 &  7.99 $\times$ 4.50 &   9.78 \\ 
2019-09-02 &   3.45 &        263 &        476 &    1.57 &  3308 &  8.51 $\times$ 4.34 &  -3.24 \\ 
2019-09-03 &   3.45 &        263 &        695 &    0.61 &   881 &  7.80 $\times$ 4.52 &   8.57 \\ 
2019-09-09 &   3.42 &        264 &        495 &    1.62 &  3260 &  8.25 $\times$ 4.36 &  -3.39 \\ 
2019-09-12 &   3.46 &        262 &        737 &    0.60 &   808 &  8.16 $\times$ 4.34 &   8.66 \\ 
2019-09-13* &  3.45 &        263 &        390 &    1.63 &  4183 &  7.98 $\times$ 4.53 &   9.49 \\ 
2019-09-14* &  3.46 &        262 &        417 &    1.63 &  3920 &  8.03 $\times$ 4.79 &   1.50 \\ 
2019-09-15 &   3.46 &        262 &        495 &    1.61 &  3251 &  8.07 $\times$ 4.52 &   6.32 \\ 
2019-09-16 &   3.42 &        264 &        716 &    0.59 &   821 &  7.71 $\times$ 4.80 &   9.01 \\ 
2019-09-24* &  3.35 &        267 &        377 &    1.62 &  4303 &  8.16 $\times$ 4.55 &  -1.18 \\ 
2019-09-29 &   3.45 &        263 &        727 &    1.59 &  2182 &  8.03 $\times$ 4.64 &   1.57 \\ 
\hline
\multicolumn{8}{c}{Combined images} 
\\
\hline
2018 & 10.40 & 151 & 240 & 1.37 & 5701 & 8.81 $\times$ 3.81 & 0.11 \\
2019 & 10.26 & 152 & 214 & 1.46 & 6811 & 8.81 $\times$ 3.81 & 0.11 \\
2018+2019 & 20.66 & 107 & 158 & 1.41 & 8974 & 8.81 $\times$ 3.81 & 0.11 \\
\enddata
\tablecomments{Summary of VLITE image properties observed in the A configuration.  Column 1: Observing date. The * denotes data used in the combined images. Column 2: Total on-source time.  Column 3: Theoretical sensitivity calculated using the VLA Exposure Calculator and assuming all 16 VLITE antennas are available, a central frequency of 338.55 MHz, a bandwidth of 30~MHz (to account for the typical $\sim$10\% loss due to RFI from the total pipeline processed bandwidth of 33.6 MHz), and robust weighting.  Column 4: Image rms noise. Column 5: Max pixel intensity.  Column 6: Dynamic range, defined as $S_{\rm max}$/$\sigma_{\rm image}$.  Column 7: Clean beam major $\times$ minor axes.  Column 8: Position angle of the clean beam. }
\end{deluxetable*}

\subsection{Data Combination and Final Images} 
\label{sec:combination}

To increase sensitivity, three observations from each of the 2018 and 2019 observing epochs were selected for further processing (see Table~\ref{tab:18A-199}). The 2018 datasets were chosen based on their high-quality pipeline images and to optimize $uv$ coverage. In 2019, ionospheric conditions introduced direction- and time-dependent source position shifts in many images, distorting point sources into triangular shapes. Three days with minimal ionospheric distortion were selected.

The selected datasets were re-calibrated as described in Section~\ref{sec:reduction}, using only the 3C48 scan included in each observation block as the primary calibrator. A wider bandwidth of 38.2~MHz centered at 340.9~MHz was retained, and data were visually inspected for RFI and flagged as needed. The three datasets from each year were combined and jointly imaged, with two rounds of phase-only, direction-independent self-calibration applied.

For each year, we converted the self-calibrated, $uv$-combined FITS datasets into measurement sets using the Common Astronomy Software Applications package \citep[CASA;][]{casa+22} and re-imaged them with joint-channel deconvolution in WSClean \citep{offringa+14, offringa+17}, which provided the highest quality image for each year. The resulting 2018 and 2019 images had synthesized beams of $8.6^{\prime\prime} \times 3.3^{\prime\prime}$ (BPA~$= 3^{\circ}$) and $8.2^{\prime\prime} \times 3.7^{\prime\prime}$ (BPA~$= -3^{\circ}$), with 1$\sigma$ rms noise levels of 225 and 205~$\mu$Jy~bm$^{-1}$, respectively, at their native resolution.

\begin{figure*}[t!]
\centering
\includegraphics[clip=true, trim=4cm 1cm 5.5cm 3cm, width=\textwidth]{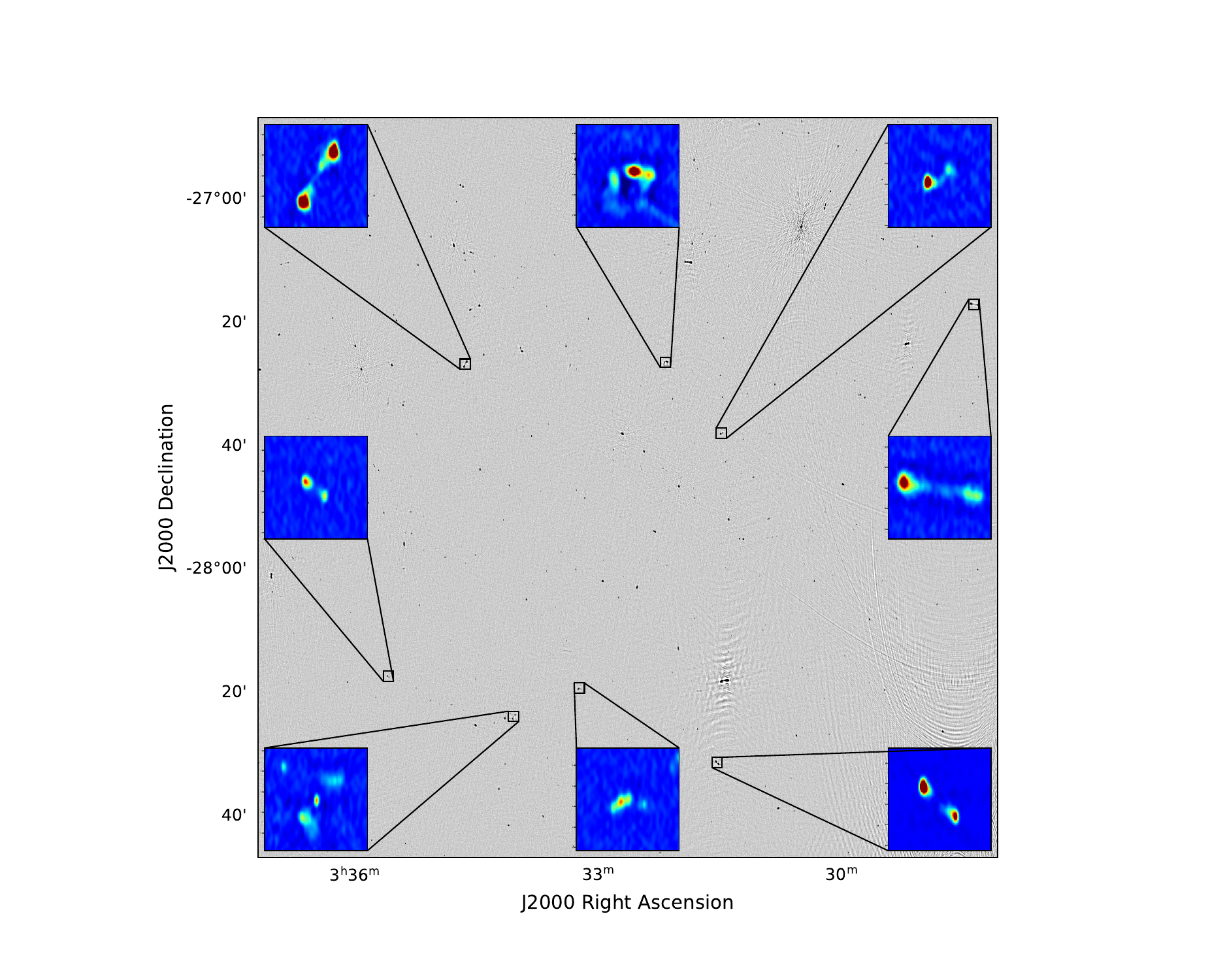}
\caption{Deep 340~MHz VLITE image centered on the HUDF field, produced from 20.66~hr of data spanning six observing days across two VLA A-configuration semesters in 2018 and 2019. The image covers $2^\circ\times2^\circ$ with a synthesized beam of $8.8^{\prime\prime} \times 3.8^{\prime\prime}$ and a 1$\sigma$ rms noise level of 158~$\mu$Jy~bm$^{-1}$. Insets highlighting examples of extended sources are $100^{\prime\prime}$ on a side.}
\label{fig:image}
\end{figure*}

In both years, VLITE recorded data from a subset of 16 of the 27 VLA antennas. The hour-angle and $uv$ ranges of the two years were nearly identical, and most antenna positions were the same. Combining the visibilities of both years for joint deconvolution would offer only a negligible improvement in $uv$ coverage. Morever, the version of WSClean available when these data were imaged could not jointly deconvolve datasets with different antenna tables, and joint deconvolution in Obit left additional artifacts around extended sources. We therefore convolved both yearly images to a common resolution of $8.8^{\prime\prime} \times 3.8^{\prime\prime}$ (BPA~$= 0^{\circ}$) and combined them in the image plane to produce the final deep image shown in Figure~\ref{fig:image}.

\begin{figure}[ht]
    \centering
    \includegraphics[width=\columnwidth]{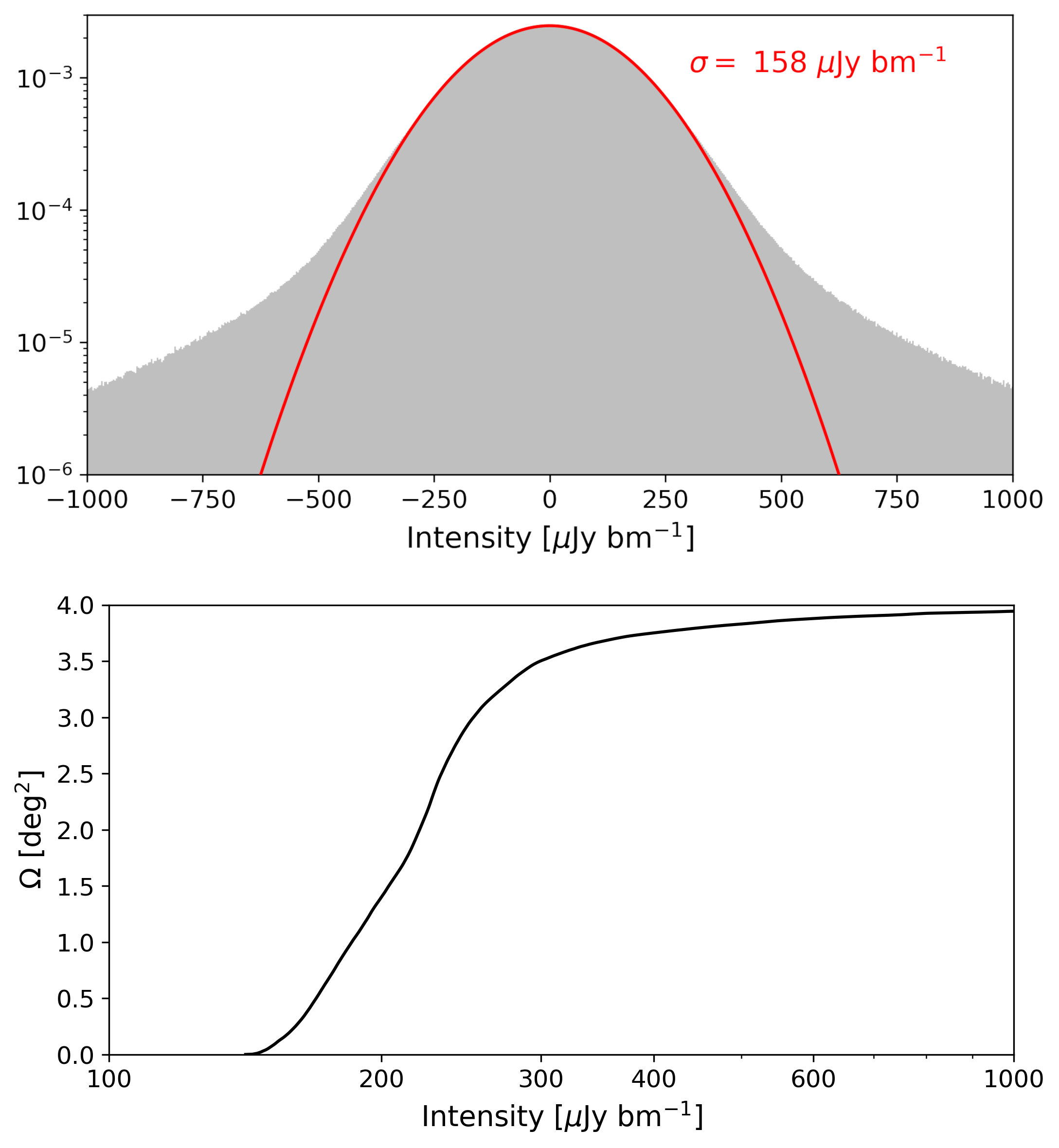}
    \caption{\textit{Top:} Histogram of pixel intensities in the final deep image after source subtraction. The distribution is predominantly Gaussian with $\sigma = 158$~$\mu$Jy~bm$^{-1}$, but exhibits a non-Gaussian tail at $|I| \gtrsim 3\sigma$, reflecting residual calibration errors and sidelobe confusion. \textit{Bottom:} Cumulative solid angle as a function of local noise level in the final image, after primary beam correction. Half the field area has a local noise below 220~$\mu$Jy~bm$^{-1}$ and 90\% is below 330~$\mu$Jy~bm$^{-1}$.}
    \label{fig:noise}
\end{figure}

The quality of the final combined image is characterized in Figure~\ref{fig:noise}. Figure~\ref{fig:noise} (top) shows the normalized pixel intensity distribution of the source-subtracted residual image prior to primary beam correction. The distribution is well described by a Gaussian component with $\sigma = 158$~$\mu$Jy~bm$^{-1}$, along with a non-Gaussian tail that dominates at intensities beyond $3\sigma$. Such non-Gaussian noise is expected for sparse, wide-field interferometric imaging without direction-dependent calibration, and is a consistent feature of VLITE images; it arises from residual calibration errors and sidelobe confusion, and its abundance and spatial distribution vary from image to image. Figure~\ref{fig:noise} (bottom) shows the cumulative solid angle as a function of local noise level after primary beam correction. Half of the field area has a local noise below 220~$\mu$Jy~bm$^{-1}$, and 90\% of the field is below 330~$\mu$Jy~bm$^{-1}$.

\section{HUDF Source Catalog}
\label{sec:catalog}

We present a catalog of 671 radio sources detected at $\geq 5\sigma$ significance in the final deep 340~MHz image. Source finding and flux density measurements were performed using the VLITE Database Pipeline \citep[VDP,][]{vdp}, a collection of Python scripts that automates source extraction and characterization in VLITE images. VDP integrates the Python Blob Detector and Source Finder \citep[PyBDSF,][]{pybdsf} for source detection. The following subsections describe the source finding configuration, primary beam calibration, compactness classification, flux density calibration, astrometric correction, and the Gaussian components comprising multi-component sources.

The full catalog is described in Appendix~\ref{sec:catalog_columns}, which provides column definitions (Table~\ref{tab:catalog}) and is released in its entirety as a machine-readable table. The fitted Gaussian components of the multi-component sources are similarly described in Appendix~\ref{app:gaussians} (Table~\ref{tab:gaussians}).

\subsection{Source Finding Configuration}

Source extraction was performed using PyBDSF, which identifies sources by fitting elliptical Gaussians to pixel peaks within islands of contiguous pixels. Island boundaries and locations for initializing Gaussian fitting were defined using the default thresholds of \texttt{thresh\_isl}~$= 3$ and \texttt{thresh\_pix}~$= 5$, corresponding to three and five times the local noise, respectively, after subtraction of the mean background. PyBDSF was configured to calculate the mean background and local noise images using a sliding 2D box with interpolation. Based on prior experience with VLITE images, we set the \texttt{rms\_box} parameter to one-tenth of the image size and the sliding step to one-third the box size. 

To improve the sampling of the local noise estimate near bright sources, we enabled an adaptive refinement grid by setting \texttt{adaptive\_rms\_box}~$=$~True, defined bright sources as those with S/N~$>$~10 (\texttt{adaptive\_thresh}~$= 10$), and set \texttt{rms\_box\_bright} to one-fifth of the nominal \texttt{rms\_box}. We lowered \texttt{flag\_maxsize\_fwhm} to 0.2 to mitigate cases of bright sources incorrectly flagged as too compact. This configuration minimized the identification of bright artifacts as genuine emission while retaining sensitivity to real sources.

PyBDSF was executed in the same manner for all daily pipeline and combined images, adopting default parameter values unless specified above. The initial run on the final deep image detected 680 sources, of which 14 were identified as spurious detections or artifacts upon visual inspection, yielding a preliminary catalog of 666 sources.

To ensure catalog completeness, we ran PyBDSF iteratively on residual images until no additional sources with S/N $>5$ were detected. In the first residual image, 72 candidate sources were identified, of which visual inspection confirmed five as genuine. Subsequent iterations uncovered 20 additional artifacts but no new genuine sources. The final catalog contains 671 validated sources.

\subsection{Primary Beam Calibration}
\label{sec:pbcor}

VLITE receivers are mounted at the prime focus of an off-axis Cassegrain optical path, which offsets the primary beam center by $6.5^{\prime}$ from the antenna pointing direction. Because the VLA antennas employ alt-azimuth mounts, the primary beam pattern also rotates with changing parallactic angle throughout each observation. Together these effects produce a smeared, arc-shaped attenuation pattern across the field of view, punctuated by gaps due to RFI excision and interruptions during antenna slews to calibrator sources. As a result, the effective primary beam response is distinct for each VLITE image.

We model the VLITE primary beam response following \citet{poli2024}, who describe a method for computing the smeared beam pattern corresponding to each VLA subreflector position. These beam models are computed and applied directly to the daily pipeline images by VDP. For the combined images, we compute a weighted average of the daily primary beam attenuations, where the weights reflect the fractional contribution of visibilities from each day to the final image. The estimated uncertainty of the primary beam calibration is $<3$\% \citep{poli2024}.

\subsection{Source Compactness}
\label{sec:compactness}

To identify unresolved sources, we adopt a well-established method based on a compactness metric that combines the flux density-to-brightness ratio $R = S_t/S_p$, where $S_t$ is the integrated (total) flux density and $S_p$ is the peak flux density, together with S/N \citep[e.g.,][]{tgss, degas2018, frail2024}. This approach accounts for the correlation between measurement uncertainties in flux density and brightness with S/N, which arises from biases inherent to source-finding algorithms \citep{hales2012, hop2015}. We define the compactness metric as the ratio of the expected value of $R$, derived from a fitted boundary function in the $R$--S/N parameter space, to the measured value. Sources with a compactness metric greater than unity have a $\sim$97\% probability of being unresolved.

We determine the boundary function with a statistical fitting procedure applied to single-component sources (\texttt{S\_Code}~$=$~`S'; see Appendix~\ref{sec:catalog_columns}), binned by S/N. Extended sources are excluded by iteratively removing $2\sigma$ outliers within each bin and computing the mean and standard deviation of $R$. The upper $2\sigma$ envelope of the distribution is then fit using an empirical equation generalized from \citet{will2013}:
\begin{equation}
    S_t/S_p = a_0 + \sqrt{a_1 + c_0\,(S/N)^{c_1}}
\end{equation}
This functional form is advantageous because $a_0$ and $a_1$ can be estimated directly from the median flux ratio and standard deviation of high-S/N sources where limited source counts prevent reliable bin statistics. The remaining parameters, $c_0$ and $c_1$, are determined by fitting the upper $2\sigma$ values in bins at low and intermediate S/N, where larger source counts enable robust bin statistics.

\begin{figure}[ht]
    \centering
    \includegraphics[width=\columnwidth]{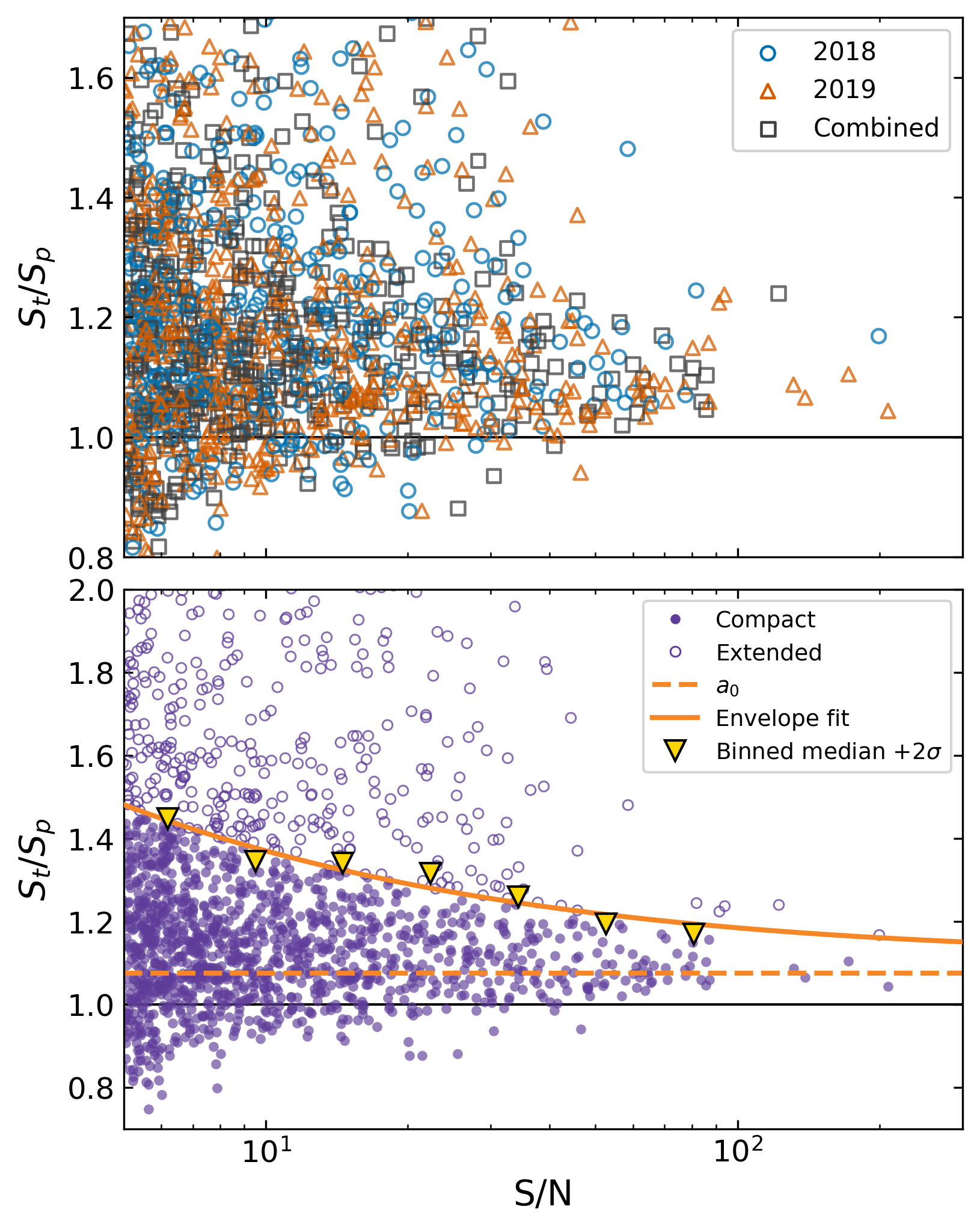}
    \caption{Flux density ratio as a function of S/N for single-component sources in the 2018, 2019, and final combined images. \textit{Top}: The three samples plotted separately (2018, blue circles; 2019, orange triangles; combined, gray squares), showing close agreement between them. The vertical range is truncated to highlight the bulk of the distribution. \textit{Bottom}: All three samples together. Yellow triangles show the binned upper $2\sigma$ values used to fit the boundary function (solid line) that statistically separates compact sources (filled circles) from extended sources (open circles). The dashed line indicates the asymptotic flux ratio at high S/N, corresponding to the fitted parameter $a_0$.}
    \label{fig:frat}
\end{figure}

Figure~\ref{fig:frat} shows the flux ratio as a function of S/N for single-component sources in the 2018, 2019, and final combined images. The top panel shows the three samples separately. Their flux ratio distributions are in close agreement, with median $S_t/S_p$ values at fixed S/N that differ by $<2\%$. Although ionospheric conditions were generally poorer in 2019, the three days combined for the 2019 image were selected for minimal ionospheric distortion (Section~\ref{sec:combination}). Accordingly, the 2019 sample does not show elevated flux ratios; if anything, its median ratio is slightly lower than that of 2018, indicating no measurable additional smearing in the 2019 image.

Given this agreement, we combine the three samples to determine the boundary function (Figure~\ref{fig:frat}, bottom). Including the 2018 and 2019 images increases the number of measurements per bin, although the three samples are not independent. The data were divided into seven logarithmic bins spanning S/N values from 5 to 100. The parameters $a_0$ and $a_1$ were determined from the median and standard deviation of sources with S/N~$>$~60, after iteratively removing $2\sigma$ outliers. The best-fit parameter values are $a_0 = 1.0761$, $a_1 = 0.0022$, $c_0 = 0.7395$, and $c_1 = -0.9429$. The compact source distribution asymptotes to a constant offset $a_0 > 1$ at high S/N and shows an additional upward trend at low S/N. The physical origins of both features are discussed in Section~\ref{sec:calib}.

\subsection{Flux Density Uncertainty and Calibration Accuracy}
\label{sec:calib}

To establish the absolute flux density scale accuracy of the final deep image, we first characterize the flux density uncertainty of the daily pipeline images from which the final image is derived. We construct a sample of point sources detected in the daily images within $7^{\prime\prime}$ of a final catalog position, where the matched final catalog source is non-multicomponent and compact (compactness $>1$), and with more than 5 detections in the daily images. The final sample comprises 131 sources with a total of $3,184$ detections across all 45 daily images.

\begin{figure}[ht]
    \centering
    \includegraphics[width=\columnwidth]{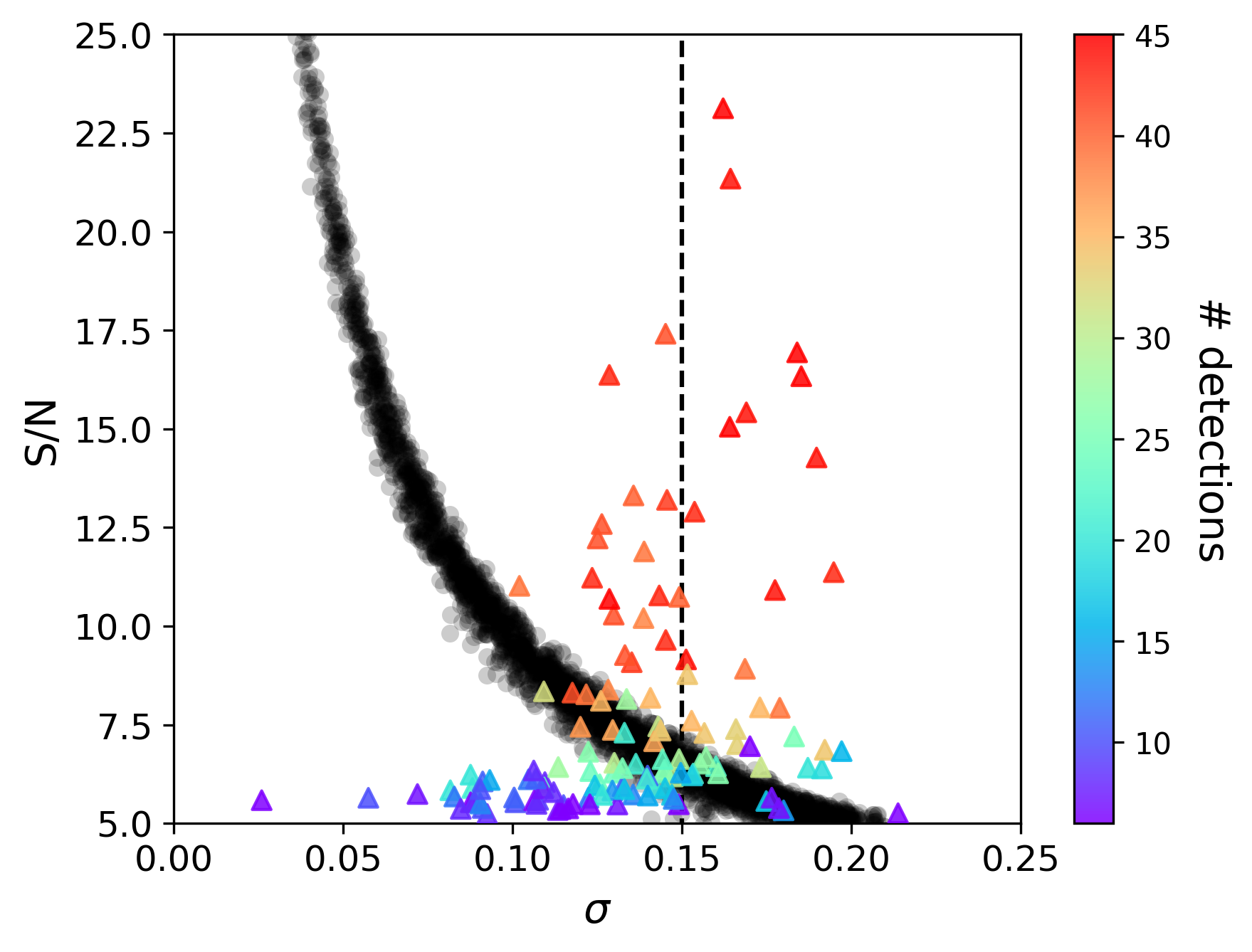}
    \caption{Comparison of VLITE flux density uncertainties as a function of S/N. Triangles show the normalized light curve standard deviation ($\sigma_{S_p}/\overline{S_p}$) for a sample of 131 point sources in the daily images, color-coded by number of detections. Gray points show PyBDSF fitted $S_p$ normalized uncertainties for individual detections. Fitting uncertainties follow the expected $\sigma = (S/N)^{-1}$ relation, while light curve scatter is approximately constant at $\sim$15\% (dashed line), confirming that calibration uncertainty dominates over thermal noise for bright sources.}
    \label{fig:sigma}
\end{figure}

Figure~\ref{fig:sigma} confirms that calibration uncertainty, rather than thermal noise, dominates the flux density uncertainty in the daily images. We characterize each source light curve by its median S/N across all daily detections. PyBDSF fitting uncertainties of individual detections decrease with S/N following $\sigma = (S/N)^{-1}$ \citep{condon1997}, as expected for thermal-noise-limited measurements. In contrast, the normalized scatter in source light curves is $\sim$15\% and roughly constant with median S/N, consistent with the calibration accuracy expected for VLITE daily images (Section~\ref{sec:reduction}) and indicating a systematic noise floor set by calibration rather than by source brightness.

At low median S/N, source light curves fall systematically below the $\sim$15\% calibration floor. This reflects a selection bias: faint sources are preferentially detected on days when positive noise fluctuations boost their apparent flux above the 5$\sigma$ threshold, causing their light curve scatter to be underestimated relative to the true variability (Eddington bias).

The daily pipeline images, though individually accurate only to $\sim$15\%, are calibrated against \citet{perley+17} independently each day. For well-sampled sources, the resulting scatter should converge toward the true \citet{perley+17} flux density over many independent daily measurements. Comparing the daily images catalog to the final catalog therefore provides an indirect check on the absolute flux calibration of the final deep image, in place of the standard approach of comparison against an independent catalog at a similar frequency, which is not available for the HUDF field.

\begin{figure}[ht]
    \centering
    \includegraphics[width=\columnwidth]{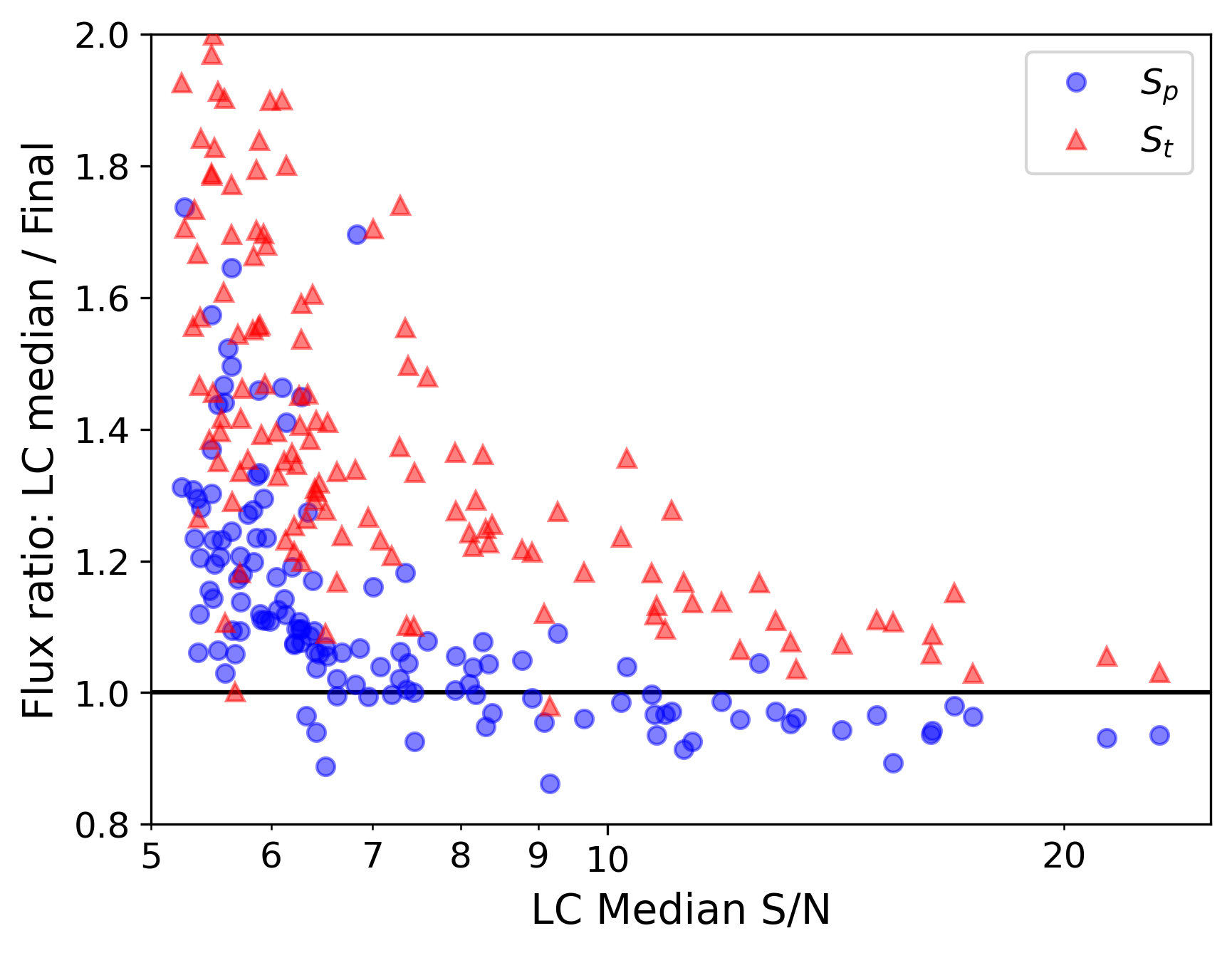}
    \caption{Ratio of daily pipeline light curve median flux densities to final deep image flux densities for 131 compact, single-component catalog sources, as a function of median daily S/N. Blue circles show peak flux density ratios ($S_p^{\rm daily}/S_p^{\rm final}$) and red triangles show total flux density ratios ($S_t^{\rm daily}/S_t^{\rm final}$). The sharp upturn at S/N $\lesssim 7$ reflects Eddington bias, while the steady rise in $S_t$ at intermediate S/N is driven by source-fitting overestimation bias exacerbated by non-Gaussian VLITE image noise.}
    \label{fig:day2final}
\end{figure}

Figure~\ref{fig:day2final} shows the ratio of daily median light curve to final image flux densities for the same source sample. Daily $S_t$ values show a steady rise at intermediate S/N driven by a known overestimation bias in source fitting algorithms \citep{hales2012, hop2015}, which is exacerbated by the non-Gaussian noise properties of VLITE images. A sharp upturn at median S/N $\lesssim 7$ is attributable to the same Eddington bias noted above. At high S/N, $S_t^{\rm daily}/S_t^{\rm final}$ approaches unity from above, with the highest S/N sources showing a residual offset of $\sim$3\%, consistent with the known source-fitting overestimation bias approaching zero asymptotically rather than indicating a true flux scale offset.

In contrast, $S_p^{\rm daily}/S_p^{\rm final}$ shows no intermediate-S/N rise but converges to $\sim$0.94 at high S/N: daily peak flux densities are $\sim$6\% lower than in the final image while total flux densities agree. This is the signature of source broadening, which suppresses $S_p$ while conserving $S_t$. Among the daily images, peak flux densities decrease relative to the final image as $\sigma_{\rm image}/\sigma_{\rm thermal}$ increases. Total flux densities show no such trend, as expected from residual phase errors, including those introduced by the ionosphere. The days selected for the final image (Section~\ref{sec:combination}) have the lowest $\sigma_{\rm image}/\sigma_{\rm thermal}$ values, so the final image is less affected by broadening than the typical daily image.

Sources in the final image are nonetheless broadened relative to the restoring beam. Since $S_t = S_p \times \Omega_{\rm fit}/\Omega_{\rm beam}$, where $\Omega_{\rm fit}$ and $\Omega_{\rm beam}$ are the fitted source and restoring beam solid angles, the asymptotic ratio $a_0 \approx 1.076$ (Figure~\ref{fig:frat}) implies that fitted sizes of compact sources exceed the restoring beam even at high S/N. The excess of the fitted minor axes over the restoring beam is independent of S/N, indicating genuine broadening in the image rather than a source-fitting bias. The excess along the major axis does decrease with S/N, reflecting fitting bias, which scales with the beam size and is therefore largest along the major axis, superimposed on the same broadening. This broadening exceeds the typical level in VLITE A-configuration data, for which $S_t/S_p$ asymptotes to $\sim$1.03 and $\sim$1.02 in the 2018 and 2019 semesters, respectively, and is similar in the 2018 and 2019 HUDF images (Section~\ref{sec:compactness}). One possible contributor is the low elevation of the HUDF, which transits at only $\sim$28$^\circ$ at the VLA, increasing the ionospheric path length relative to typical VLITE targets.

In principle, $S_t/S_p > 1$ could instead reflect overestimated total flux densities. The daily-to-final comparison favors suppressed peaks: total flux densities agree to within $\sim$3\% between images with very different degrees of broadening, whereas peak flux densities vary with image quality. $S_t$ is therefore the more robust measure of the flux density scale. Because this argument is indirect, we do not apply a correction but instead adopt the $\sim$6\% peak offset as a conservative upper bound on any residual absolute flux scale uncertainty. Combining in quadrature with the $<$3\% primary beam calibration uncertainty (Section~\ref{sec:pbcor}) and the 3--5\% absolute accuracy of the \citet{perley+17} flux density scale itself at these frequencies, we adopt $\sim$8\% as the $1\sigma$ absolute flux density scale uncertainty for the final VLITE HUDF catalog.

To establish the relative flux density scale between the 2018 and 2019 combined images, which underpins the variability analysis of Section~\ref{sec:var}, we directly compare flux densities measured in the two year-combined images. We construct the sample from the final catalog, selecting compact (Section~\ref{sec:compactness}), isolated sources fit by a single Gaussian (\texttt{S\_Code}~$=$~`S') that have counterparts with the same \texttt{S\_Code} in both the 2018 and 2019 image catalogs, yielding a sample of 216 sources. Restricting the sample in this way avoids flux density differences arising from blending with neighboring sources or from the greater sensitivity of extended source fits to differences in image noise between the two epochs.

\begin{figure}[ht]
    \centering
    \includegraphics[width=\columnwidth]{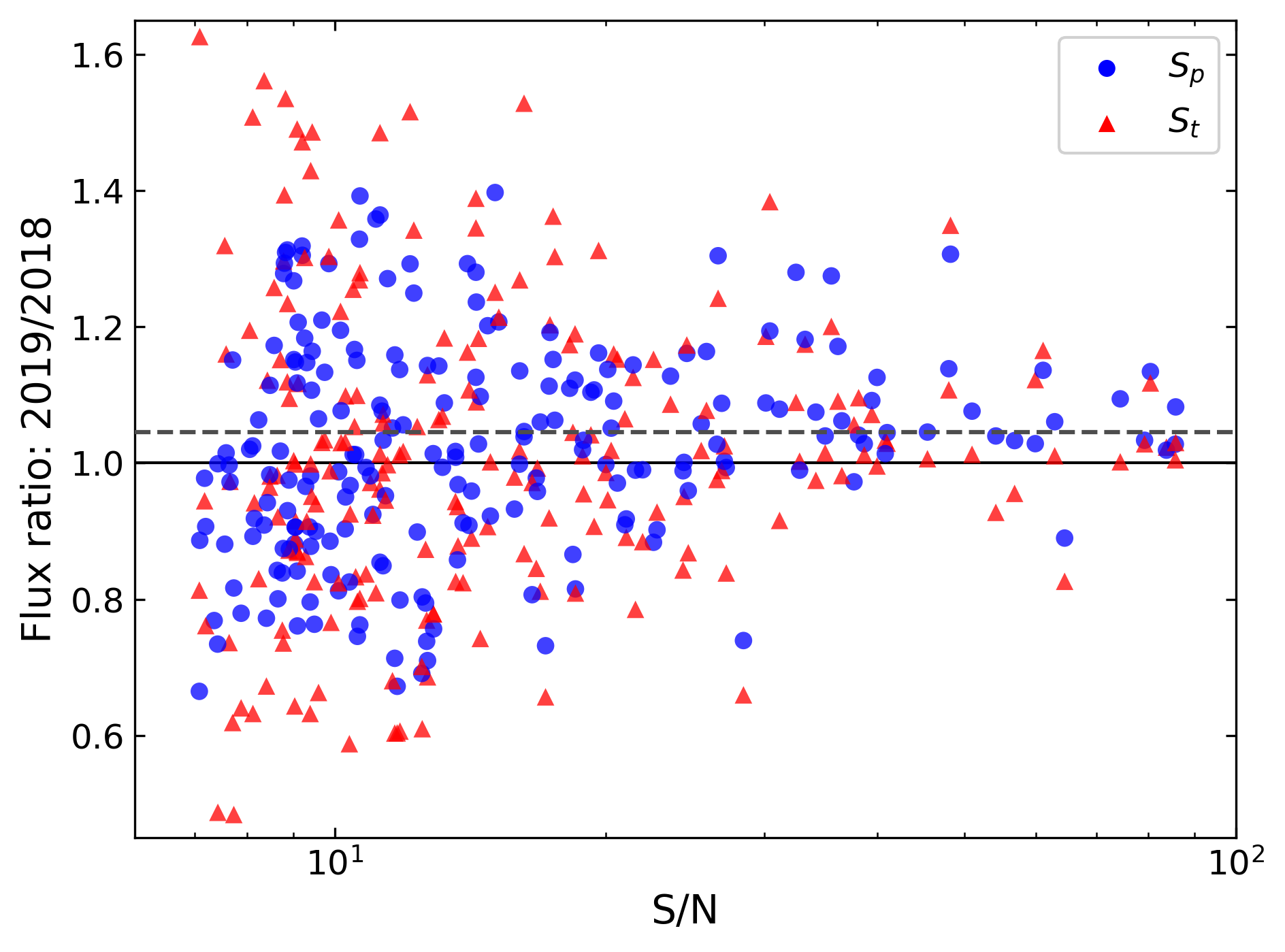}
    \caption{Ratio of 2019 to 2018 flux densities for 216 compact, single-component sources detected in both year-combined images, as a function of S/N. Flux densities are measured in the 2018 and 2019 images; the final catalog is used only to cross-identify the sources and to provide the S/N values. Blue circles show peak flux density ratios ($S_p^{\rm 2019}/S_p^{\rm 2018}$) and red triangles show total flux density ratios ($S_t^{\rm 2019}/S_t^{\rm 2018}$). Both ratios converge to the same offset of $\sim$5\% at high S/N (dashed line).}
    \label{fig:yr2yr}
\end{figure}

Figure~\ref{fig:yr2yr} shows the ratio of 2019 to 2018 flux densities as a function of source S/N. Both $S_p$ and $S_t$ show a stable offset of $\sim$5\% across all S/N, indicating a calibration-level difference between the two epochs rather than S/N-dependent fitting effects. For a steady source, a $\sim$5\% flux offset would produce an apparent modulation index of only $\sim$2.5\%. This is more than a factor of two below the smallest modulation index among the variable sources identified in Section~\ref{sec:var}, confirming that residual year-to-year calibration differences do not contaminate the variable sample.

\subsection{Astrometric Accuracy}
\label{sec:astrometry}

To evaluate astrometric errors in source positions, we cross-match the VLITE source catalog with the ATLAS HUDF catalog at 1.4~GHz \citep{atlas1400}. ATLAS is well suited for this purpose due to its large overlap area of 2.7~deg$^2$, comparable angular resolution of $16.3^{\prime\prime} \times 6.8^{\prime\prime}$, and deep sensitivity of 14~$\mu$Jy~beam$^{-1}$, sufficient to detect all VLITE sources in our sample. We identify 282 compact, single-component VLITE sources matched within $7^{\prime\prime}$ of an ATLAS counterpart. Of these, 101 are bright sources with VLITE S/N~$>$~10, where fitting errors are expected to be $<$10\% \citep{condon1997}. We use this bright sample to characterize and correct the VLITE astrometry.

\begin{figure}[htb!]
    \centering
    \includegraphics[width=\columnwidth]{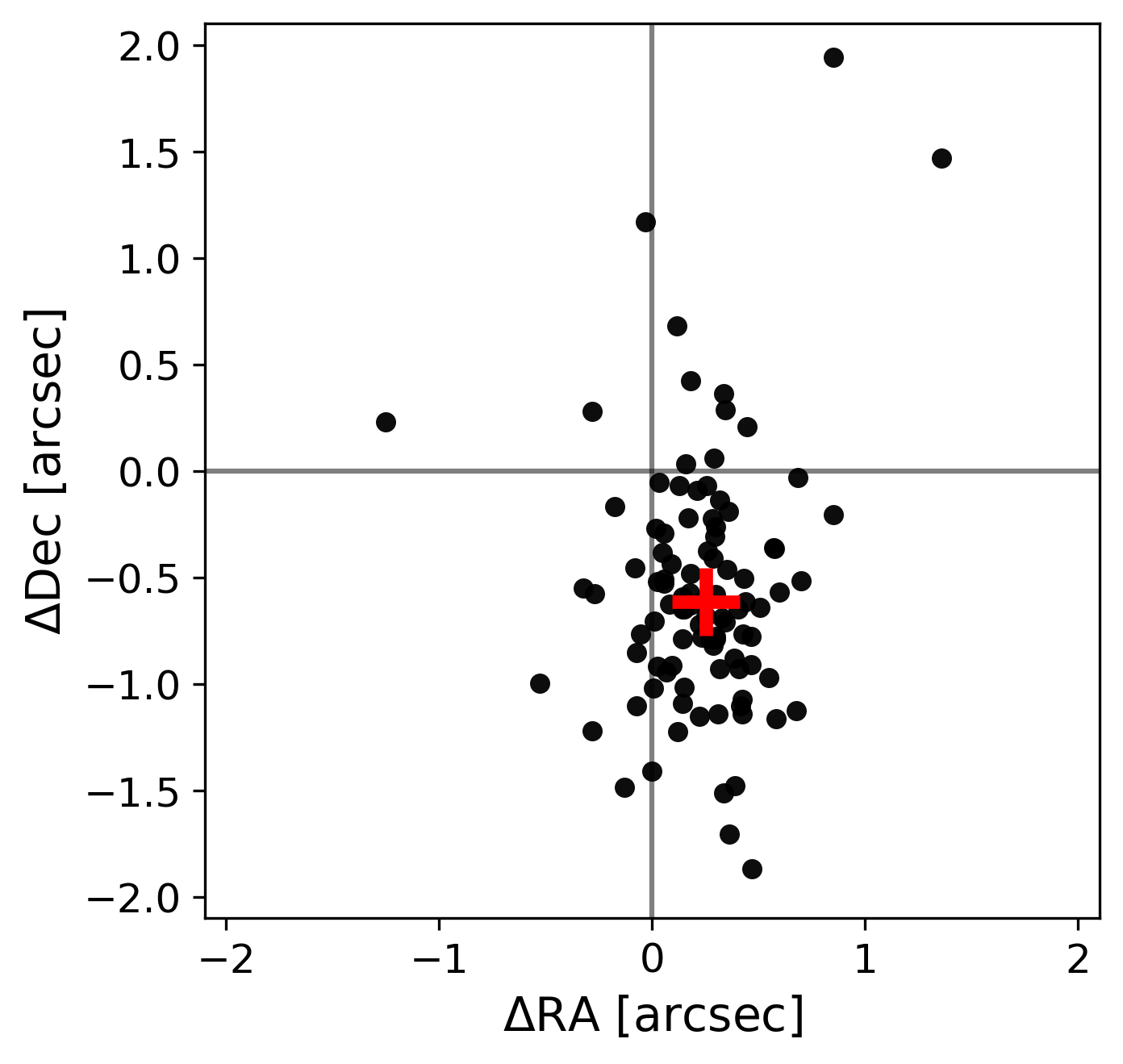}
    \caption{Positional offsets in right ascension and declination between 101 bright VLITE sources and their ATLAS 1.4~GHz counterparts, prior to astrometric correction. The cross marks the median offset in each 
    coordinate.}
    \label{fig:off}
\end{figure}

Figure~\ref{fig:off} shows the coordinate offsets for the bright source sample. The median offsets and standard deviations are $0.25 \pm 0.32^{\prime\prime}$ in RA and $-0.62 \pm 0.83^{\prime\prime}$ in Dec., indicating a small but systematic positional offset, particularly in declination. We apply these corrections to all VLITE source coordinates and adopt $1^{\prime\prime}$ as the $1\sigma$ systematic positional uncertainty.

\subsection{Gaussian Component Catalog}
\label{sec:gauscomp}

Of the 671 sources in the final catalog, 121 (18\%) are classified by \texttt{PyBDSF} as multi-component ($\texttt{S\_Code} = $`M'), together comprising 363 individual fitted Gaussian components (out of 913 Gaussian components fitted across the full catalog). For single-Gaussian sources ($\texttt{S\_Code} = $`S' or `C'), the fitted Gaussian and the cataloged source are identical, and Table~\ref{tab:catalog} fully describes them. For multi-component sources, however, the total flux density reported in Table~\ref{tab:catalog} is the sum of multiple, spatially distinct Gaussian components, and reconstructing this sub-structure requires the individual component parameters. We release the fitted parameters of all 363 Gaussian components belonging to the 121 multi-component sources in Appendix~\ref{app:gaussians} (Table~\ref{tab:gaussians}).

\section{Analysis}
\label{sec:analysis}

With the final catalog of radio sources in hand, we examine the properties of the detected population. We assess the completeness of the catalog through injection-recovery simulations, evaluate the spectral index distribution by cross-matching with a higher-frequency survey, and characterize a population of variable sources identified by comparing flux densities across the 2018 and 2019 VLITE epochs.

\subsection{Completeness and Source Counts}
\label{sec:completeness}

Variations in noise across the field of view introduce brightness-dependent incompleteness in the VLITE HUDF catalog. In particular, sources with sufficient integrated flux but extended over large angular scales can be missed because their peak brightness falls below the detection threshold. To quantify catalog completeness as a function of source flux density, we performed injection-recovery simulations using modeled source populations.

We adopted model populations from the Tiered Radio Extragalactic Continuum Simulation \citep[T-RECS,][]{trecs}, which characterizes the AGN and SFG populations of the radio sky from 150~MHz to 20~GHz. We used the published medium-tier catalog covering a $5^{\circ} \times 5^{\circ}$ region at 300~MHz, scaling flux densities downward by 8.4\% to match VLITE's 340~MHz observing frequency. For steep-spectrum AGN (FR~I/II types; T-RECS PopFlag~$= 6$), we multiplied catalog sizes by the $R_s$ parameter to approximate the projected hotspot separation, which characterizes the extent of most radio emission, and assumed circular symmetry. For SFGs, sizes were scaled by 0.7 to convert to the exponential disk scale radius, with ellipticity applied to derive the minor axis.

Model sources were convolved with the VLITE restoring beam, weighted by the primary beam response at each source position, and injected into a residual image from which all sources and artifacts $>5\sigma$ had been subtracted. We generated five mock images, four covering the corners of the T-RECS field and one centered field that partially overlaps the others, to sample a range of noise environments. Each mock image was processed with PyBDSF using identical parameters to the real data, and primary beam corrections were applied to the recovered flux densities. Sources were binned by flux density and the detection fraction was averaged across the five fields. Table~\ref{tab:comp} summarizes the results.

\begin{deluxetable}{lrrrr}
\tablecaption{Completeness simulation results\label{tab:comp}}
\tablehead{
\colhead{Flux bin (mJy)} & \colhead{$N_{\rm inj}$} & \colhead{$N_{\rm det}$} & \colhead{$N_{\rm obs}$} & \colhead{$C$}
}
\startdata
0.54--0.93   & 753.6 & 13.2  & 15  & 57.1 \\
0.93--1.30   & 251.6 & 54.8  & 63  & 4.6  \\
1.3--1.8     & 158.8 & 98.4  & 104 & 1.6  \\
1.8--3.0     & 156.2 & 149.0 & 152 & 1.0  \\
3--4         & 61.2  & 62.6  & 59  & 1.0  \\
4--8         & 97.2  & 97.6  & 103 & 1.0  \\
8--16        & 66.0  & 65.2  & 70  & 1.0  \\
16--32       & 36.4  & 37.2  & 31  & 1.0  \\
32--64       & 21.6  & 22.0  & 30  & 1.0  \\
64--150      & 15.8  & 15.6  & 22  & 1.0  \\
150--300     & 5.4   & 5.4   & 14  & 1.0  \\
300--600     & 5.4   & 5.4   & 6   & 1.0  \\
\enddata
\tablecomments{$N_{\rm inj}$: mean number of sources injected per mock image; $N_{\rm det}$: mean number recovered; $N_{\rm obs}$: number observed in the real image; $C = N_{\rm inj}/N_{\rm det}$: completeness correction factor.}
\end{deluxetable}

\begin{figure}[ht]
    \centering
    \includegraphics[width=\columnwidth]{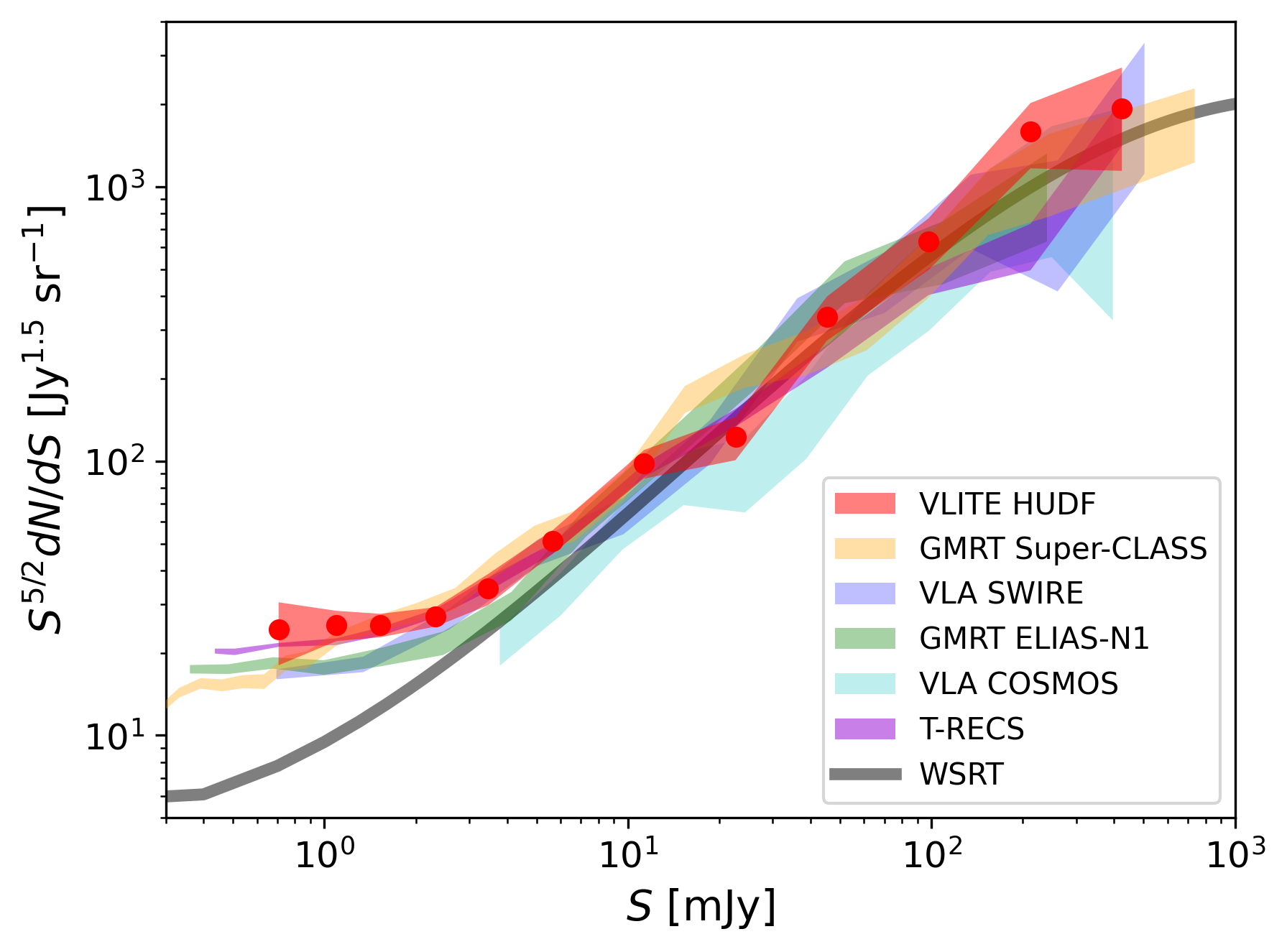}
    \caption{Corrected VLITE HUDF source counts (red) compared to other deep surveys at similar frequencies (Table~\ref{tab:Pband_surveys}) and the T-RECS model catalog. The black line shows the WSRT fit from \citet{wieringa} (Eqn.~\ref{eq:wsrt}). Shaded areas indicate $1\sigma$ Poisson uncertainties only and do not include the additional field-to-field sample variance discussed in the text. Completeness corrections have been applied to the three faintest VLITE flux bins.}
    \label{fig:counts}
\end{figure}

Figure~\ref{fig:counts} shows the corrected VLITE HUDF source counts, with completeness corrections from Table~\ref{tab:comp} applied to the three faintest flux bins. We compare the corrected counts to other deep surveys at similar frequencies (Table~\ref{tab:Pband_surveys}) and to the T-RECS model catalog used to derive the corrections. For reference, the black line shows a fit to 327~MHz source counts from the Westerbork Synthesis Radio Telescope \citep[WSRT,][]{wieringa}:
\begin{multline}
    \log_{10}(S^{5/2}\,dN/dS) = 0.976 + 0.6136x \\ 
    + 0.3028x^2 - 0.083x^3
    \label{eq:wsrt}
\end{multline}
where $x = \log_{10}(S)$ in mJy. This fit applies to sources brighter than 4~mJy and does not capture the rise in source counts at lower fluxes due to the increasing contribution of SFGs.

The VLITE counts are consistent with the T-RECS model at the faint end. This agreement is not guaranteed by construction: the completeness correction (Table~\ref{tab:comp}) rescales the observed counts by the T-RECS recovery fraction rather than by the model's predicted source density, so a true HUDF population departing from T-RECS would show up as a departure in the corrected counts rather than being masked by the correction. The agreement therefore indicates that T-RECS provides a reasonable description of the sub-mJy population in this field, modulo any residual dependence on the source size and structure assumptions used to estimate completeness.

As an independent check, we compare the HUDF counts to three independent P-band fields with sub-mJy sensitivity and comparable-to-larger area: GMRT ELIAS-N1 \citep[][3.8~deg$^2$]{sirothia+09}, VLA SWIRE \citep[][3.1~deg$^2$]{owen+09}, and GMRT Super-CLASS \citep[][6.5~deg$^2$]{riseley+16}. The HUDF counts differ by no more than 1.6$\sigma$ at any flux density, once field-to-field sample variance \citep[][scaled to each survey's solid angle]{heywood+13} is combined in quadrature with Poisson uncertainty. The sign of this offset varies between comparison fields rather than showing a consistent excess, as expected for sample variance rather than a systematic effect. The small solid angle of the HUDF field makes source counts sensitive to large-scale structure along the line of sight, and field-to-field variations of this magnitude are expected \citep[e.g.,][]{heywood+13}.

\subsection{Spectral Indices}
\label{sec:spix}

We calculate spectral indices as a consistency check on the measured flux densities in our catalog. For the compact, non-variable sources that dominate our sample, we expect spectral indices broadly consistent with synchrotron-emitting AGN and SFGs. Systematic deviations would indicate calibration errors, flux scale offsets, or source matching problems.

Spectral indices were computed for the 282 compact, single-component sources matched to the ATLAS 1.4~GHz catalog, as described in Section~\ref{sec:astrometry}. The ATLAS flux densities are referenced to the \citet{baars1977} absolute scale with an estimated uncertainty of 2\%. VLITE flux densities are calibrated to the \citet{perley+17} scale with an estimated accuracy of $\sim$8\%. Since the two scales agree to within 1\% at 328~MHz \citep{perley+17}, no correction is applied to the ATLAS flux densities.

\begin{figure*}[ht!]
    \centering
    \includegraphics[width=\textwidth]{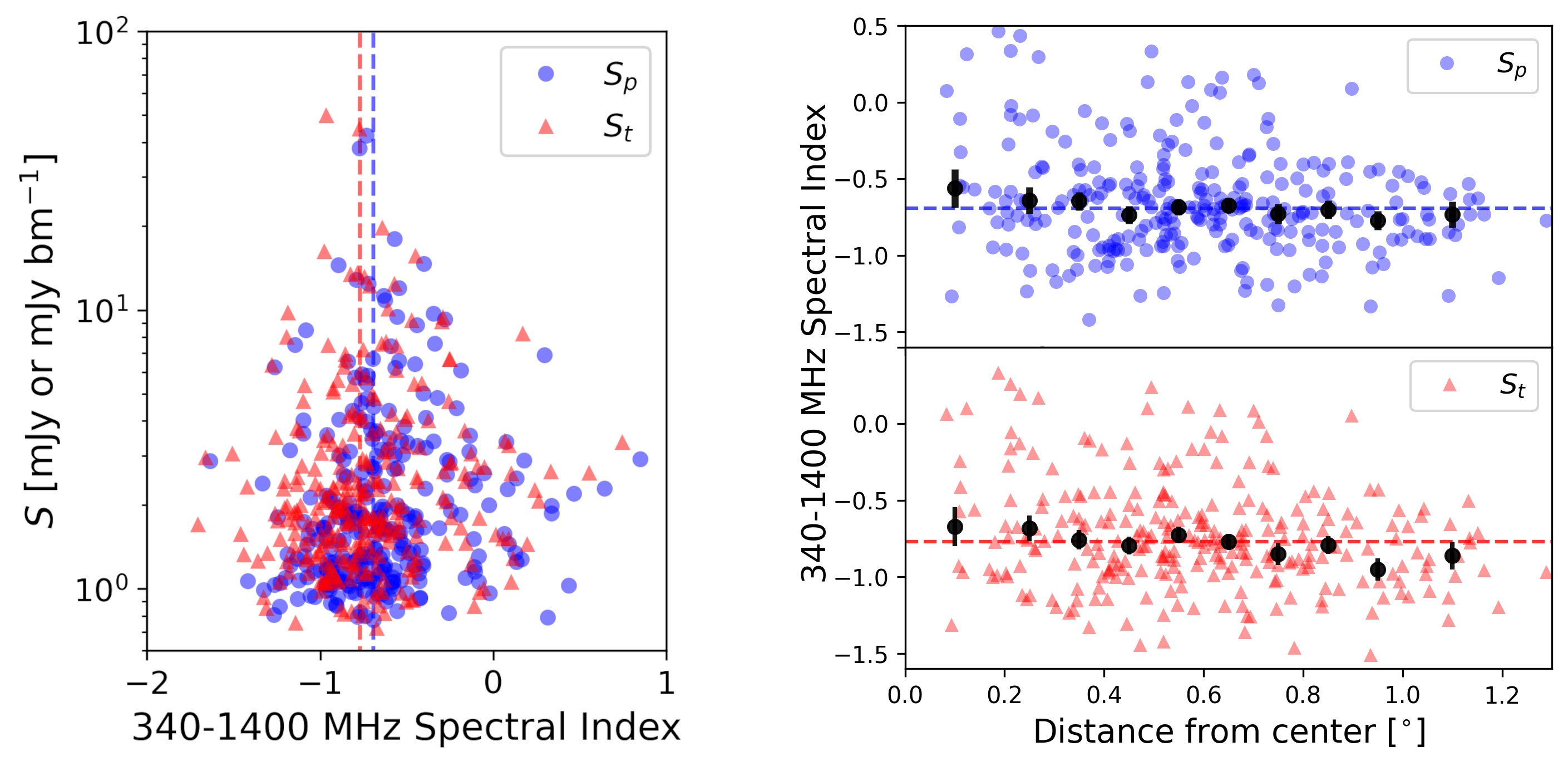}
    \caption{\textit{Left:} Spectral index flux density distribution for compact, single-component sources matched to the ATLAS 1.4~GHz catalog. Blue circles denote spectral indices computed using peak flux densities ($S_p$), while red triangles correspond to total flux densities ($S_t$). Dashed lines indicate the median values: $-0.69$ for $S_p$ and $-0.77$ for $S_t$. \textit{Right:} The top and bottom panels show spectral indices computed using peak and total flux densities, respectively, as a function of angular distance from the VLITE pointing center. Dashed lines indicate the median values across the full sample. Black points show median spectral indices in radial bins, with error bars indicating the standard error of the median.}
    \label{fig:spix}
\end{figure*}

The left panel of Figure~\ref{fig:spix} presents the spectral index distribution as a function of VLITE flux density. The median spectral indices are $-0.69^{+0.30}_{-0.27}$ and $-0.77^{+0.33}_{-0.30}$ for peak and total flux densities, respectively, where the uncertainties give the 16th--84th percentile range of the distribution. The difference between the two medians reflects the flux density biases discussed in Section~\ref{sec:calib}: correcting $S_p$ for the $\sim$7\% systematic suppression from source broadening steepens the median peak spectral index to $-0.74$, while the higher median $S_t$ index reflects the S/N-dependent upward fitting bias near the detection threshold. All values are consistent with expectations for optically thin synchrotron emission \citep{condon+92} and with previous P-band studies, which report typical spectral indices in the range $-0.70$ to $-0.83$ over similar frequency intervals \citep{owen+09, sirothia+09, smolcic+14, coppejans2015, mauch2013, riseley+16}.

The right panel of Figure~\ref{fig:spix} shows the spectral index as a function of angular distance from the VLITE pointing center. Median spectral indices are plotted in ten distance bins spanning $0^\circ$--$1.2^\circ$, with error bars reflecting the standard error of the median in each bin. Larger uncertainties at small ($<0.2^\circ$) and large ($>1^\circ$) distances reflect reduced source counts in those regions. Across most of the field the bin medians deviate from the overall population median by no more than 0.07, confirming that the primary beam calibration introduces no significant systematic spectral gradient.

\subsection{Variable Sources}
\label{sec:var}

One of VLITE's primary scientific objectives is the identification of transient and variable sources of low-frequency radio emission. The multi-epoch HUDF dataset provides a direct measure of VLITE's sensitivity to variability on $\sim$1-year timescales at sub-mJy flux densities.

Our analysis focuses on a sample of 389 single-component sources from the final catalog, each detected with S/N~$>$~5 in either the 2018 or 2019 combined images. Following \citet{mooley2016}, we define a two-epoch variability metric $V = |\Delta S|/\sigma$, where $\Delta S$ is the flux density difference between epochs and $\sigma$ is the quadrature sum of the fitted flux density uncertainties. For sources with S/N~$<$~5 in either epoch, we used PyBDSF's force-fitting functionality to recover flux measurements. We evaluated the variability metric separately using $S_p$ and $S_t$ and classified candidate variable sources as those for which both $V_p$ and $V_t$ exceeded 3.

\begin{figure}[ht]
    \centering
    \includegraphics[width=\columnwidth]{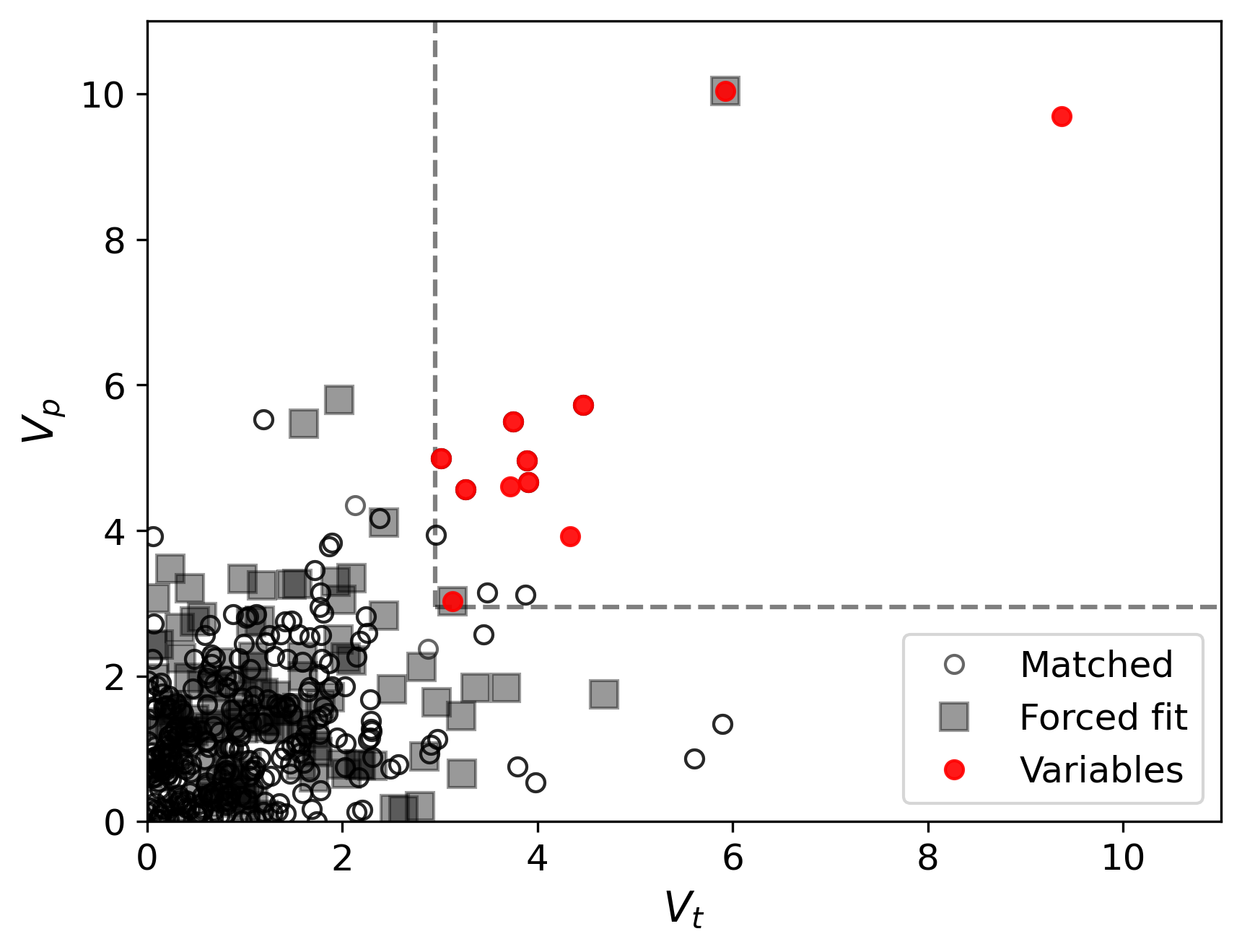}
    \caption{Variability metrics computed from peak ($V_p$) and total ($V_t$) flux densities for single-component sources detected in the combined 2018 and 2019 epoch images. Open circles indicate sources with S/N $> 5$ in both epochs, while squares denote sources for which flux densities were obtained through forced fitting in one epoch. Sources within the boxed region, exhibiting high values of both $V_p$ and $V_t$, were classified as candidate variables. Filled circles mark eleven sources confirmed as genuine variables following additional validation.}
    \label{fig:var}
\end{figure}

Figure ~\ref{fig:var} presents the distribution of variability metrics for the full source sample. Eleven single-component sources met the candidate threshold. We further evaluated these candidates through visual inspection and comparison with flux densities measured in the daily pipeline images, and produced difference images by subtracting the 2018 and 2019 epoch images to independently validate the variability. Sources with high variability metrics exhibited prominent residuals in the difference images, consistent with genuine variability.

During this visual inspection, we identified three additional bright sources with prominent residuals in the difference images that had not been included among the original candidates because PyBDSF classified them as multi-component. Closer examination revealed that this classification arose from weak noise features within the source island fitted as additional Gaussians alongside the primary component. In each case the primary Gaussian accounts for more than 95\% of the grouped flux density, and the sources are more accurately described as single-component. Computing the variability metrics using only the primary Gaussian flux densities, we confirmed that both $V_p$ and $V_t$ exceed the candidate threshold in each case, bringing the total candidate sample to fourteen sources as shown in Figure~\ref{fig:var}.

Of these fourteen candidates, one was determined to be spurious, with elevated variability arising from a noise artifact. Two others showed significantly weaker variability in their daily light curves than in the two-epoch comparison, and were rejected. We thus identify a final sample of eleven sources as robust variables, as indicated in the boxed region of Figure~\ref{fig:var}.

Figure~\ref{fig:figBIGa} presents a summary of the properties of the eleven variable sources. For each source, the left panels show the 340~MHz light curves from the 2018 and 2019 epochs, with the combined-year flux densities indicated by the shaded bands. Flux density measurements from the daily pipeline images are shown as filled symbols where PyBDSF detected the source at S/N~$\geq 5$, and as open symbols where force-fitting was used to recover measurements for weaker detections with S/N~$\approx 4$--5. Three sources have no daily image measurements in one or both epochs: J0335$-$2649 falls in a corner of the square field of view that lacks coverage in the daily images, while the other two are too faint to be recovered even with force-fitting. 

\begin{figure*}
    \centering
    \includegraphics[width=\textwidth]{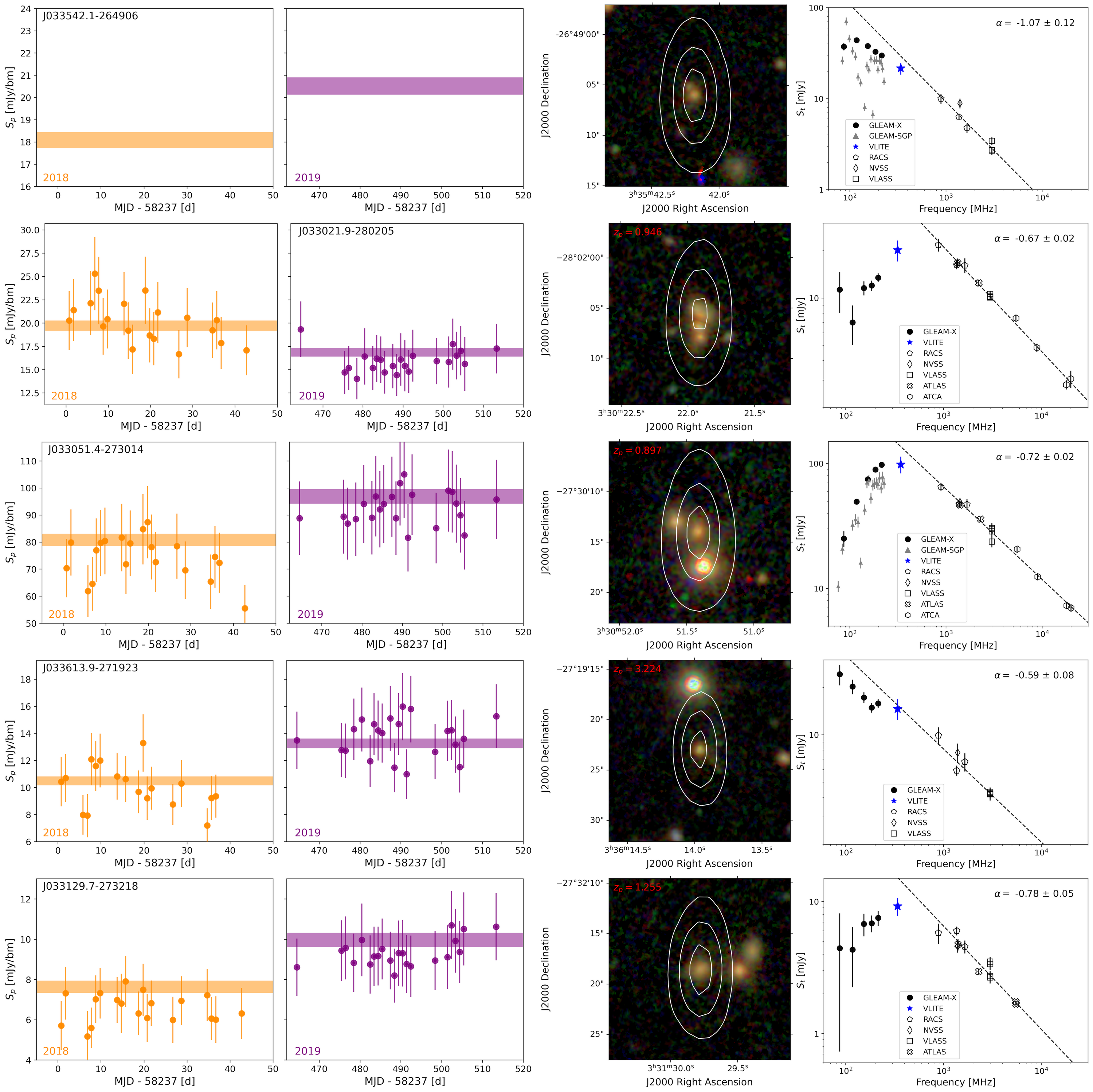}
    \captcont{Eleven sources exhibiting variability on $\sim 1$~yr timescales. {\it Left two panels:} Light curves from the 2018 and 2019 epochs; shaded bars represent intensities measured from the combined year images. {\it Middle-right panels:} VLITE radio contours overlain on infrared images from the {\it Euclid} $Y$, $J$, and $H$ bands. {\it Right panels:} Radio spectra for each source assembled from the surveys listed in Table~\ref{tab:var}. Data $> 800$~MHz (open symbols) were fitted to a power-law model (dashed lines). Sources are ordered top to bottom by increasing modulation index.}
    \label{fig:figBIGa}
\end{figure*}

\begin{figure*}
    \centering
    \includegraphics[width=\textwidth]{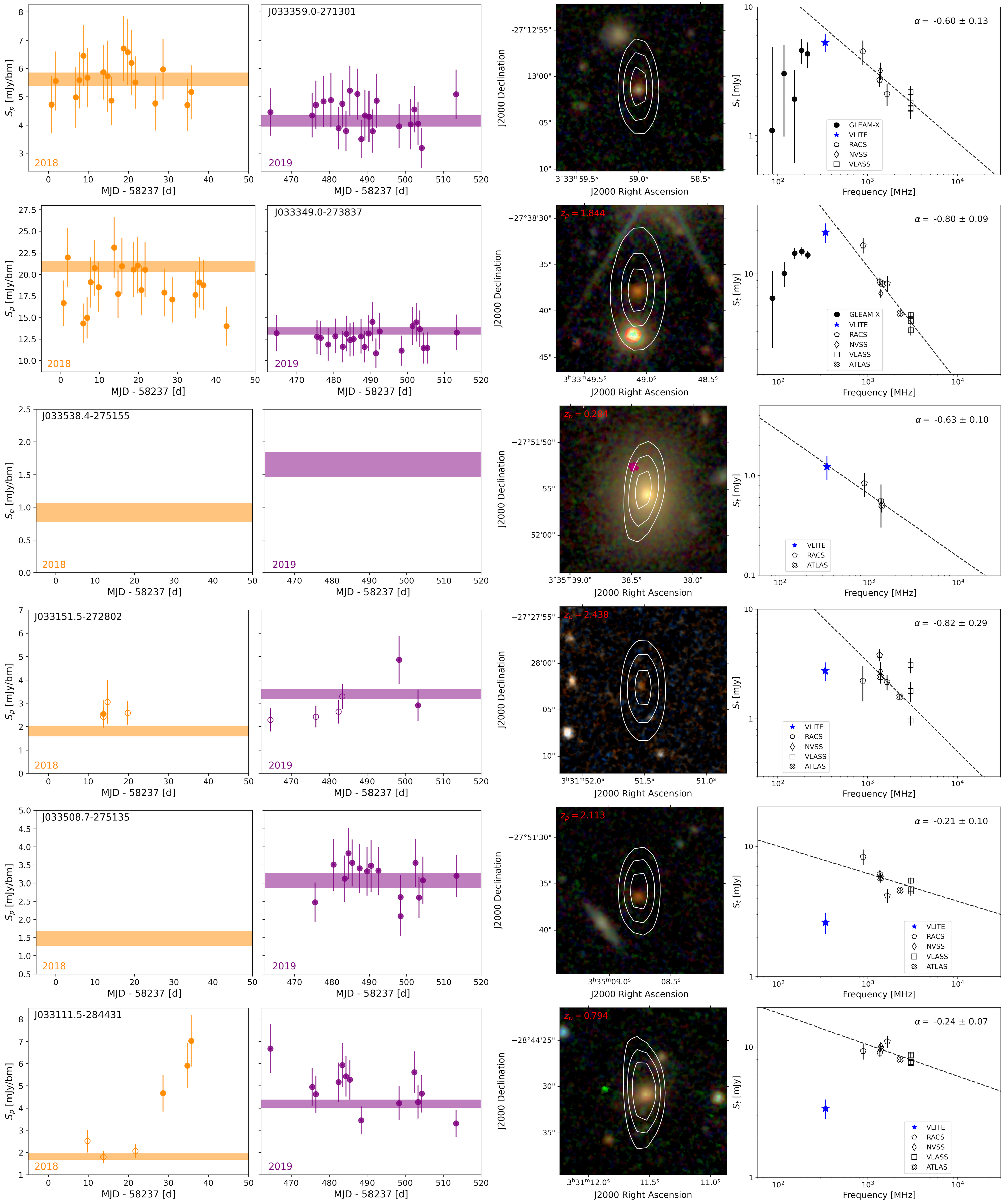}
    \caption{\it Continued.}
\end{figure*}

The middle panels of Figure~\ref{fig:figBIGa} show three-color infrared images from the {\it Euclid} $Y$, $J$, and $H$ bands \citep{euclid+26} with VLITE radio contours overlaid. The right panels show the broadband radio spectra assembled from survey measurements spanning MHz to GHz frequencies, with a power-law fit to the optically thin data above 800~MHz. Sources are ordered from top to bottom by increasing modulation index, defined as half the difference in peak flux density between the two epochs relative to their mean. 

Table~\ref{tab:var} summarizes the key properties of each variable source, including the two-epoch variability metrics, modulation index, photometric redshift from DeepDrill \citep{deepdrill,nyland+23}, and spectral classification. For sources with measured photometric redshifts, we compute the 1.4~GHz radio luminosity following \citet{condon+18}:
\begin{equation}
L_\nu = 4\pi D_L^2 S_\nu (1+z)^{-\alpha - 1}
\label{eq:luminosity}
\end{equation}
where $D_L$ is the luminosity distance, $S_\nu$ is the fitted flux density at 1.4~GHz, and $\alpha$ is the spectral index. We adopt a flat $\Lambda$CDM cosmology with $H_0 = 67.7$~km~s$^{-1}$~Mpc$^{-1}$ and $\Omega_m = 0.307$ \citep{planck+16}. For the two sources lacking photometric redshifts we assume $z = 1$, consistent with the typical redshift of radio-loud AGN at these flux densities; the resulting luminosities are marked with a dagger ($^\dagger$) in Table~\ref{tab:var}.

\begin{deluxetable}{lccccc}
\tablecaption{Properties of the eleven sources exhibiting variability on $\sim$1~yr timescales.\label{tab:var}}
\tablewidth{\columnwidth}
\tablehead{
\colhead{Name} & 
\colhead{($V_t$, $V_p$)} & 
\colhead{$m_p$} & 
\colhead{$z_{\rm p}$} & 
\colhead{$L_{\rm 1.4~GHz}$} & 
\colhead{Spectral} \\
\colhead{} & 
\colhead{} & 
\colhead{(\%)} & 
\colhead{} & 
\colhead{(W Hz$^{-1}$)} & 
\colhead{Class} \\
\colhead{(1)} & 
\colhead{(2)} & 
\colhead{(3)} & 
\colhead{(4)} & 
\colhead{(5)} & 
\colhead{(6)} 
}
\startdata
J0335$-$2649     & (4.6, 3.3) & 6  & \nodata & 3.7E+25$^\dagger$ & PS \\
J0330$-$2802     & (4.3, 3.9) & 8  & 0.946  & 6.5E+25   & PS \\
J0330$-$2730     & (3.7, 4.6) & 9  & 0.897  & 1.7E+26   & PS \\
J0336$-$2719     & (4.5, 5.7) & 12 & 3.224  & 3.5E+26   & PL \\
J0331$-$2732     & (3.9, 5.0) & 13 & 1.255  & 4.1E+25   & PS \\
J0333$-$2713     & (3.9, 4.7) & 15 & \nodata & 1.2E+25$^\dagger$ & PS \\
J0333$-$2738     & (9.4, 9.7) & 22 & 1.844  & 1.7E+26   & PS \\
J0335$-$2751$^a$ & (3.0, 3.1) & 28 & 0.284  & 1.3E+23   & PL \\
J0331$-$2728     & (5.0, 3.0) & 31 & 2.438  & 9.8E+25   & FS \\
J0335$-$2751$^b$ & (3.8, 5.5) & 35 & 2.113  & 8.1E+25   & FS \\
J0331$-$2844     & (5.9, 10.0)& 40 & 0.794  & 1.9E+25   & FS \\
\enddata
\tablecomments{Columns: (1) Source name, truncated to HHMM$-$DDMM; names marked $^a$ and $^b$ share the same truncated designation but refer to distinct sources  identifiable by their full names in Figure~\ref{fig:figBIGa}. (2) Variability metrics from total and peak flux; (3) Modulation index from peak flux; (4) Photometric redshift from \citet{nyland+23}; (5) 1.4 GHz luminosity; $^\dagger$Luminosity calculated assuming $z = 1$. (6) Source classification. Radio spectra shown in Figure~\ref{fig:figBIGa} were assembled from the following surveys: GLEAM-X 72--231~MHz \citep{gleamx}; GLEAM-SGP 72--231~MHz \citep{gleamsgp}; RACS-low 887.5~MHz \citep{racslo}; RACS-mid 1367.5~MHz \citep{racsmiddr2}; RACS-high 1655.5~MHz \citep{racshigh}; NVSS 1.4~GHz \citep{condon+98}; VLASS 3~GHz \citep{vlassql}; ATLAS 1.4~GHz \citep{atlas1400}; ATCA 5.5, 9, 18, and 20~GHz \citep{franzen+14}.}
\end{deluxetable}

The eleven variable sources share several properties that constrain their physical nature. Their radio spectra fall into three broad classes: six sources with peaked spectra (PS), consistent with MHz-peaked spectrum (MPS) sources; three with flat spectra (FS); and two with simple power-law spectra (PL). The peaked-spectrum sources have spectral turnovers that place them in the same class as compact, young radio AGN whose emission is absorbed either by synchrotron self-absorption (SSA) or free-free absorption (FFA) in a dense circumnuclear medium \citep{odea+98, callingham+17}. 

The 1.4~GHz radio luminosities computed from the fitted spectra are $> 10^{25}$ W~Hz$^{-1}$, consistent with radio-loud AGN and quasars \citep[e.g.,][]{gold+99}. The exception is J0335$-$2751$^a$, whose luminosity of $1.3 \times 10^{23}$~W~Hz$^{-1}$ is at the low-luminosity end of radio-loud AGN. The {\it Euclid} infrared images show that all sources are associated with nuclear emission from elliptical host galaxies, consistent with the AGN interpretation. 

\subsubsection{RISS and the Low-frequency Variable Population}
\label{sec:riss}

At low radio frequencies, the dominant extrinsic variability mechanism for compact sources is RISS \citep{rickett86, rickett90}. RISS arises when radio waves from a compact source are scattered by electron density inhomogeneities in the turbulent interstellar medium (ISM). The HUDF lies at a high Galactic latitude of $b = -54.4\degr$, where the electron density and scattering strength are significantly reduced relative to lines of sight closer to the plane. Reduced scattering lowers the transition frequency $\nu_0$ between the strong and weak scattering regimes relative to lower-latitude lines of sight. Since the modulation index in the strong scattering regime scales as $m \propto (\nu/\nu_0)^{17/30}$, a lower $\nu_0$ counterintuitively produces \textit{larger} predicted variability amplitudes at 340~MHz despite the weaker overall scattering. However, reduced scattering also increases the diffractive scale $r_{\rm diff}$, which sets the angular size of the scattering disk as $\theta_{\rm scat} \approx \lambda / (2\pi r_{\rm diff})$. A larger $r_{\rm diff}$ therefore produces a smaller scattering disk, requiring source components to be more compact in order to scintillate, since only emission regions smaller than $\theta_{\rm scat}$ contribute to the RISS modulation.

For a continuously and densely sampled light curve, the RISS variability amplitude is characterized by the modulation index $m = \sigma_S / \bar{S}$, where $\sigma_S$ is the standard deviation of the flux density and $\bar{S}$ is the mean. With only two epochs available, we approximate this as $m_p = 0.5|S_{p,1} - S_{p,2}|/\bar{S_p}$, which reduces to the population standard deviation divided by the mean for a two-point sample. We note that this definition differs from alternatives in the literature by factors of $\sqrt{2}$--2, introducing comparable uncertainty in comparisons with RISS models, though this is subdominant to the limitation of characterizing a stochastic process from only two measurements.

We use the \texttt{RISS19} software \citep{riss19}, which computes RISS properties self-consistently from H$\alpha$ intensity maps of the ionized ISM \citep{finkbeiner+03}, to estimate the scattering parameters for the HUDF line of sight. For comparison, the analytic estimates of \citet{walker98} give consistent results: \texttt{RISS19} predicts $\nu_0 \approx 4$--$6$~GHz for this field, consistent with the Walker analytic estimate of $\nu_0 \approx 6$~GHz, confirming that our 340~MHz observations are well within the strong scattering regime. The predicted point-source modulation index is $m \approx 20\%$ at 340~MHz, and the scattering disk size is $\theta_{scat} \approx 2$~mas, independent of screen distance for extragalactic sources, corresponding to a physical scale of $\sim$16~pc at $z \sim 1$.

The predicted refractive timescale depends on the assumed screen distance $D$ and effective transverse velocity $v_{eff}$ through the scaling $t_r \propto D^{1/2} v_{eff}^{-1}$ \citep{walker98}. The daily pipeline images provide an observational constraint on these parameters: for the lower-modulation-index sources in our sample,  the daily light curves within each epoch show no evidence of complete variation cycles over timescales of $\sim$40~days, implying $t_r \gtrsim 0.1$~yr. This places an upper limit on the effective screen velocity of $v_{eff} \lesssim 80$~km~s$^{-1}$ at $D = 1.0$~kpc, decreasing to $v_{eff} \lesssim 30$~km~s$^{-1}$ at $D = 0.3$~kpc, consistent with typical effective ISM velocities at high Galactic latitudes. Across this parameter range $\nu_0$ varies from $\approx$ 4.1--5.8~GHz and the predicted modulation index remains robustly $m \approx 20$--$24\%$ throughout.

The predicted 24\% maximum RISS modulation is consistent with the observed variability for the seven sources with $m_p \lesssim 22\%$. However, RISS requires that the radio-emitting region be smaller than the scattering disk ($\theta_{\rm scat} \approx 2$~mas, or $\sim$16~pc at $z \sim 1$). The turnover--size relation \citep{odea+97} for MPS sources predicts lobe extents of order 1--2~kpc for sources peaking near 300--600~MHz, at which scale the modulation index would be suppressed by a factor of $>$100 relative to a point source \citep{keim+19}. RISS therefore requires a compact sub-component, such as a hotspot or compact double structure, that is optically thin at 340~MHz and falls within the scattering disk, even if the source is extended on larger scales \citep{ross+21}.

This picture is supported by the VLBA follow-up of our brightest variable source, J0330$-$2730 ($m_p = 9\%$), presented in \citet{nyland+26}. VLBA C-band imaging resolved J0330$-$2730 into a symmetric double-lobed structure with a projected linear size of 64~pc, classifying it as a compact symmetric object (CSO). At $\sim$8~mas angular separation the lobes are comparable in size to the scattering disk, consistent with RISS contributing to the observed variability. The brightness temperatures ($\log T_B/\mathrm{K} = 6.6$--$8.4$) indicate nonthermal synchrotron emission with no evidence for relativistic beaming. \citet{nyland+26} ruled out Doppler boosting and concluded that RISS and variable FFA are both plausible explanations for the observed 9\% modulation.

The three sources with the highest modulation indices, J0331$-$2728 ($m_p = 31\%$), J0335$-$2751$^b$ ($m_p = 35\%$), and J0331$-$2844 ($m_p = 40\%$), present a different picture. Their variability amplitudes exceed the maximum modulation predicted by RISS and are difficult to reconcile with pure scintillation of even a maximally compact source. All three have flat radio spectra, and all show evidence of variability in archival GHz flux densities from survey catalogs at different epochs, suggesting their variability is not confined to low frequencies. J0331$-$2844 additionally shows a rapid flux density increase of $\sim$300\% over $\sim$2~weeks in the 2018 daily light curve, a timescale inconsistent with the predicted $t_r \gtrsim 0.1$~yr at 340~MHz. We therefore consider intrinsic variability to be the more likely explanation for these sources, arising from mechanisms such as shocks along a jet \citep{marscher+gear85}, variable free-free absorption \citep{tingay+dekool03, tingay+15}, or changes in the accretion rate \citep{nyland+20}.

The remaining source, J0335$-$2751$^a$ ($m_p = 28\%$), occupies an intermediate position. Its modulation index modestly exceeds the RISS prediction, but given its low flux density ($S \sim 1$~mJy) the measurement uncertainties are substantial and the true modulation index may be consistent with RISS. Its radio spectrum is poorly constrained but consistent with either a power-law or a low-frequency peaked spectrum. We consider both RISS and intrinsic variability plausible pending deeper multifrequency observations.

Definitively distinguishing between RISS and intrinsic variability requires simultaneous multifrequency observations. In the strong scattering regime, the RISS modulation index increases with frequency as $m \propto (\nu/\nu_0)^{17/30}$, reaching a maximum near $\nu_0$ before decreasing above it. Using \texttt{RISS19} with the constrained screen parameters, the predicted point-source modulation rises to $\sim$50\% at 1.4~GHz and $\sim$80\% at 3~GHz. However, the refractive timescale decreases steeply as $t_r \propto (\nu_0/\nu)^{11/5}$, falling to $t_r \gtrsim 1.5$~days at 1.4~GHz and $t_r \gtrsim 7$~hr at 3~GHz. Existing multi-epoch surveys such as NVSS, FIRST, and VLASS have cadences of years to decades, far exceeding these timescales, so individual epochs cannot resolve the timing or coherence structure of rapid RISS fluctuations. Definitively testing the RISS timescale prediction would require dedicated monitoring at 1--3~GHz with a cadence of hours to days and simultaneous 340~MHz observations.

The predicted GHz amplitude increase above additionally assumes the scintillating component remains unresolved relative to the scattering disk, whose angular size shrinks steeply with frequency as $\theta_{\rm scat} \propto \nu^{-11/5}$. At 340~MHz, $\theta_{\rm scat} \approx 2$~mas, but by 3~GHz this has fallen to $\sim$17~$\mu$as, roughly two orders of magnitude smaller. A component that is only marginally compact enough to scintillate at 340~MHz would be grossly over-resolved at 3~GHz, suppressing rather than enhancing the RISS amplitude despite the frequency scaling above. The predicted increase in modulation with frequency is therefore an upper limit that applies only to components remaining compact at all frequencies considered.

We tested this expectation directly using archival VLASS 3~GHz photometry \citep{gordon+21}, available across four epochs spanning eight years for ten of the eleven variable sources (J0335$-$2751$^a$ is too faint for a VLASS detection). Although the 32-month epoch spacing cannot resolve the sub-day RISS timescale at 3~GHz, the variance across several widely-separated epochs remains a valid if noisy estimator of the total modulation amplitude. None of the ten sources show the $\sim$3.4$\times$ enhancement predicted for an unresolved point source; most instead show comparable or reduced modulation relative to 340~MHz. For J0330$-$2730, whose 8~mas VLBA size is known precisely, this can be checked quantitatively: propagating the same size through $\theta_{\rm scat}(3\,{\rm GHz})$ predicts a suppressed modulation of order 0.2\%, consistent with the near-null VLASS result for this source ($m_p \lesssim$ a few percent, consistent with zero within the photometric floor). This provides direct empirical support for the size-dependent suppression described above. The absence of enhanced GHz variability across the sample is therefore consistent with, rather than in tension with, a RISS origin for the peaked-spectrum sources. Intrinsic variability, by contrast, typically propagates from high to low frequencies with a time delay \citep{bignall+15} and can produce spectral shape changes that RISS cannot \citep{ross+21}, providing the most direct observational test for distinguishing between the two mechanisms.

\subsubsection{Implications for VLITE Variability Searches}

The detection of eleven variable sources in 4~deg$^2$, corresponding to a surface density of $\sim 2.75$~deg$^{-2}$ (or 1.6\% of the catalog sample), places a useful lower bound on the population of low-frequency variable AGN accessible to VLITE. This rate is broadly consistent with the $\sim 1.5\%$ variable fraction found by \citet{ross+21} among compact GLEAM sources at 100--230~MHz on $\sim 1$~yr timescales, and with the enhanced variability fraction seen among peaked-spectrum sources specifically \citep{bell+19, ross+21}. Together these results support the picture that peaked-spectrum sources constitute the dominant low-frequency variable population at mJy-level flux densities.

We can rule out explosive transients, such as off-axis gamma-ray burst afterglows or radio supernovae, as an explanation for the sample. All eleven sources have catalog counterparts at higher frequencies spanning decades, inconsistent with a transient origin, and are coincident with the nuclei of elliptical host galaxies rather than the star-forming environments in which such transients typically occur. This is consistent with blind low-frequency transient searches \citep[e.g.][]{murphy+17b}, which predict an order of magnitude fewer such events per deg$^2$.

A complementary population of compact nuclear radio transients with GHz-peaked spectra has been identified through joint VLA and VLITE observations by \citet{chen+25}, with one source subsequently characterized in detail by \citet{chen+26}. These sources were selected as VLASS detections with no counterparts in earlier surveys and simultaneous VLITE non-detections confirming inverted spectra, they are interpreted as young AGN jets on parsec scales associated with galaxy nuclei. Their spectral peaks above $\sim$1--6~GHz place them at an earlier evolutionary stage than the MHz-peaked sources that dominate our variable sample, and their inverted low-frequency spectra would render them undetectable or marginal in VLITE imaging. The contrast between these two populations illustrates how low- and high-frequency variability searches probe different stages of compact radio source evolution.

This contrast also highlights a broader strength of the VLITE commensal observing strategy for variability science. Although individual VLITE flux measurements carry a calibration uncertainty of $\sim 15\%$, averaging over $N$ daily images reduces the uncertainty on the mean flux. The $\sim$5\% flux density scale offset between the independently calibrated 2018 and 2019 combined images (Section~\ref{sec:calib}) demonstrates how reduced calibration uncertainty enables the detection of $\lesssim 10$\% variability with high significance for well-sampled sources. The VLA reconfiguration cycle of approximately 15~months naturally produces the multi-epoch cadence required for RISS and intrinsic AGN variability searches, with data accumulating commensally at no additional observing cost.

The HUDF dataset demonstrates that VLITE is capable of identifying populations of compact and potentially young radio sources, including CSO candidates, through low-frequency variability selection alone, complementing spectral shape selection at higher frequencies \citep{callingham+17, ross+21, slob+22} and high-frequency transient searches \citep{chen+25, chen+26}. \citet{nyland+26} showed that the variability criterion for CSO classification is effective at VLITE frequencies, suggesting that large-area VLITE monitoring could provide a new route to statistical samples of CSOs and MPS sources.

\section{Summary}
\label{sec:summary}

We have presented 340~MHz imaging and a radio source catalog of the HUDF using commensal data from VLITE. Our main results are as follows.

\begin{enumerate}

\item Our deepest image combines 20.66~hr of data from six selected observing days spanning two VLA A-configuration semesters in 2018 and 2019. The image covers $4$~deg$^2$ with a synthesized beam of $8.8^{\prime\prime} \times 3.8^{\prime\prime}$ and achieves a 1$\sigma$ rms noise of 158~$\mu$Jy~beam$^{-1}$ at the field center, making it one of the deepest P-band images obtained to date (Table~1).

\item We catalog 671 radio sources detected at $\geq 5\sigma$ significance. Source positions are accurate to $\sim 1^{\prime\prime}$ (1$\sigma$), and flux densities carry an overall calibration uncertainty of $\sim 8\%$. Peak flux densities of compact sources are systematically $\sim$7\% lower than their total flux densities, reflecting a broadening of sources in the deep image that conserves integrated flux; total flux densities are therefore the more robust measure of the flux density scale.

\item Injection-recovery simulations using T-RECS model populations show that the catalog is complete above $\sim 1.8$~mJy, with completeness corrections reaching a factor of $\sim 57$ at the faintest flux bin (0.54--0.93~mJy). Corrected source counts are consistent with other deep P-band surveys and with the T-RECS model, recovering the sub-mJy star-forming galaxy population at abundances consistent with both.

\item Spectral indices computed by cross-matching 282 compact single-component sources with the ATLAS 1.4~GHz catalog yield median values of $-0.69^{+0.30}_{-0.27}$ and $-0.77^{+0.33}_{-0.30}$ (16th--84th percentile ranges) from peak and total flux densities, respectively, or $-0.74$ from peak flux densities corrected for source broadening, consistent with optically thin synchrotron emission from AGN and SFGs. Peak-flux spectral indices show no significant systematic variation with distance from the pointing center, confirming the fidelity of the primary beam calibration across the field.

\item Comparing flux densities between the 2018 and 2019 combined epoch images, we identify eleven sources with robust evidence for variability on $\sim 1$~yr timescales, with modulation indices ranging from 6 to 40\%. Six of the eleven have spectral properties consistent with MHz-peaked spectrum sources, and all have infrared counterparts associated with the nuclear regions of elliptical host galaxies, identifying them as radio-loud AGN. The observed variable fraction of 1.6\% is consistent with rates found in other low-frequency variability surveys \citep{ross+21, bell+19}.

\item Using the \texttt{RISS19} model \citep{riss19}, we estimate a point-source modulation index of $\sim$20\% at 340~MHz, consistent with the observed variability amplitudes for seven of the eleven sources. The three sources with the highest modulation indices ($m_p = 31$--$40\%$), all of which have flat radio spectra, exceed the predicted RISS amplitude and are more likely intrinsically variable. The remaining source ($m_p = 28\%$) lies between these groups, and given its low flux density, both RISS and intrinsic variability remain plausible. For the RISS-consistent sources, VLBA follow-up of the brightest, J0330$-$2730, by \citet{nyland+26} resolved a 64~pc double-lobed morphology consistent with a compact symmetric object, with RISS and variable free-free absorption identified as the most plausible variability mechanisms.

\end{enumerate}

The HUDF dataset illustrates the potential of the VLITE commensal observing strategy for deep low-frequency continuum surveys and time-domain science. The multi-epoch cadence produced naturally by the VLA array configuration cycle, combined with VLITE's commensal operation, enables variability searches at 340~MHz with no dedicated observing time. Advanced imaging techniques applied to the full 136~hr VLITE dataset would reach significantly greater sensitivity than the 20~hr image presented here, and is deferred to future work.

\section*{Acknowledgements}
Basic research in Radio Astronomy at the U.S.\ Naval Research Laboratory is supported by 6.1 Base funding. Construction and installation of VLITE was supported by the NRL Sustainment Restoration and Maintenance fund. The National Radio Astronomy Observatory is a facility of the National Science Foundation operated under cooperative agreement by Associated Universities, Inc.

This scientific work uses data obtained from Inyarrimanha Ilgari Bundara / the Murchison Radio-astronomy Observatory. We acknowledge the Wajarri Yamaji People as the Traditional Owners and native title holders of the Observatory site. CSIRO’s ASKAP radio telescope is part of the Australia Telescope National Facility (https://ror.org/05qajvd42). Operation of ASKAP is funded by the Australian Government with support from the National Collaborative Research Infrastructure Strategy. ASKAP uses the resources of the Pawsey Supercomputing Research Centre. Establishment of ASKAP, Inyarrimanha Ilgari Bundara, the CSIRO Murchison Radio-astronomy Observatory and the Pawsey Supercomputing Research Centre are initiatives of the Australian Government, with support from the Government of Western Australia and the Science and Industry Endowment Fund. This paper includes archived data obtained through the CSIRO ASKAP Science Data Archive, CASDA (https://data.csiro.au).

This work has made use of the {\it Euclid} Quick Release Q1 data from the {\it Euclid} mission of the European Space Agency (ESA), 2025 \citep{doi_euclid}.

This work has made use of the {\it Spitzer} Survey of Deep Drilling Fields (DeepDrill) (Program ID 11086, P.I. Lacy) \citep{doi_dd}, and the DeepDrill Forced Photometry Catalog \citep{doi_ddfpc}.

\bibliographystyle{aasjournal}
\bibliography{hudf}

\appendix

\section{Catalog Data Columns and Sample}
\label{sec:catalog_columns}

We present the VLITE HUDF source catalog as a machine-readable table containing 671 sources. Table~\ref{tab:catalog} shows a representative sample of the full catalog, spanning the flux density and S/N range of the survey and all three source-structure classifications, for guidance regarding form and content. The tabulated coordinates include the astrometric correction described in Section~\ref{sec:astrometry}, and the tabulated flux densities include the primary beam correction described in Section~\ref{sec:pbcor}. Note that the tabulated flux errors are PyBDSF fitting uncertainties only and do not include the overall $\sim$8\% VLITE flux density scale uncertainty for the final deep image (Section~\ref{sec:calib}). Similarly, the tabulated PyBDSF fitting coordinate uncertainties do not include the $1^{\prime\prime}$ systematic positional uncertainty (Section~\ref{sec:astrometry}). The catalog columns are as follows.

\begin{description}
\item[\texttt{Name}] VLITE source name
\item[\texttt{RA}] Source right ascension, in degrees
\item[\texttt{e\_RA}] $1\sigma$ fitting error on the right ascension, in arcsec
\item[\texttt{DEC}] Source declination, in degrees
\item[\texttt{e\_DEC}] $1\sigma$ fitting error on the declination, in arcsec
\item[\texttt{Total\_flux}] Fitted total, integrated Stokes I flux density of the source, in mJy
\item[\texttt{e\_Total\_flux}] $1\sigma$ fitting error on the total flux density, in mJy
\item[\texttt{Peak\_flux}] Fitted peak Stokes I flux density per beam of the source, in mJy bm$^{-1}$
\item[\texttt{e\_Peak\_flux}] $1\sigma$ fitting error on the peak flux density per beam, in mJy bm$^{-1}$
\item[\texttt{Maj}] FWHM of the source major axis, in arcsec
\item[\texttt{e\_Maj}] $1\sigma$ error on the FWHM of the major axis, in arcsec
\item[\texttt{Min}] FWHM of the source minor axis, in arcsec
\item[\texttt{e\_Min}] $1\sigma$ error on the FWHM of the minor axis, in arcsec
\item[\texttt{PA}] Position angle of the source major axis measured east of north, in degrees
\item[\texttt{e\_PA}] $1\sigma$ error on the position angle of the major axis, in degrees
\item[\texttt{Isl\_Total\_flux}] Total, integrated Stokes I flux density of the island in which the source is located, in mJy. This value is calculated from the sum of all non-masked pixels in the island with values above three times the local rms
\item[\texttt{e\_Isl\_Total\_flux}] $1\sigma$ error on the total flux density of the island in which the source is located, in mJy
\item[\texttt{Isl\_rms}] Average background rms value of the island, in mJy bm$^{-1}$
\item[\texttt{Isl\_mean}] Average background mean value of the island, in mJy bm$^{-1}$
\item[\texttt{Resid\_Isl\_rms}] Average residual background rms value of the island, in mJy bm$^{-1}$
\item[\texttt{Resid\_Isl\_mean}] Average residual background mean value of the island, in mJy bm$^{-1}$
\item[\texttt{S\_Code}] PyBDSF code defining source structure:
    \begin{description}
    \item[S] a single-Gaussian source that is the only source in the island
    \item[C] a single-Gaussian source in an island with other sources
    \item[M] a multi-Gaussian source
    \end{description}
\item[\texttt{Compactness}] Source compactness metric determined from the S/N and flux ratio (Section~\ref{sec:compactness}); values $>1$ indicate a $\sim97\%$ probability of being unresolved
\item[\texttt{SNR}] Source S/N, calculated as (\texttt{Peak\_flux} $-$ \texttt{Isl\_mean})/\texttt{Isl\_rms}
\end{description}

\begin{deluxetable*}{lcccccccccc}
\rotate
\tabletypesize{\footnotesize}
\tablewidth{0pt}
\tablecaption{VLITE HUDF Source Catalog (Sample) \label{tab:catalog}}
\tablehead{
\colhead{Name} & \colhead{R.A.} & \colhead{Decl.} & \colhead{Total Flux} & \colhead{Peak Flux} & \colhead{Maj} & \colhead{Min} & \colhead{P.A.} & \colhead{S\_Code} & \colhead{SNR} & \colhead{Compactness} \\
\colhead{(VLITE-A J)} & \colhead{(deg)} & \colhead{(deg)} & \colhead{(mJy)} & \colhead{(mJy bm$^{-1}$)} & \colhead{(arcsec)} & \colhead{(arcsec)} & \colhead{(deg)} & \colhead{} & \colhead{} & \colhead{}
}
\startdata
032836.5$-$284147 & 52.152376 & $-$28.696618 & $4603.3\pm65.2$ & $2319.6\pm32.1$ & $16.66\pm0.02$ & $5.25\pm0.01$ & $5.8\pm0.1$ & M & 828.1 & 0.565 \\
033549.1$-$274918 & 53.954771 & $-$27.821780 & $29.3\pm0.7$ & $23.6\pm0.6$ & $9.06\pm0.10$ & $4.60\pm0.03$ & $0.4\pm0.3$ & S & 122.1 & 0.914 \\
033452.8$-$275813 & 53.720216 & $-$27.970286 & $9.5\pm0.3$ & $9.3\pm0.3$ & $8.85\pm0.21$ & $3.87\pm0.04$ & $178.8\pm0.8$ & C & 56.8  & 1.143 \\
033617.9$-$283608 & 54.074779 & $-$28.602306 & $15.3\pm0.6$ & $7.6\pm0.2$ & $11.73\pm0.60$ & $4.15\pm0.09$ & $174.8\pm2.0$ & M & 26.8 & 0.626 \\
033659.7$-$280339 & 54.249109 & $-$28.060879 & $11.6\pm0.7$ & $6.4\pm0.3$ & $10.08\pm0.77$ & $5.99\pm0.32$ & $169.8\pm6.1$ & S & 16.4 & 0.753 \\
033005.4$-$265132 & 52.522660 & $-$26.858891 & $3.3\pm0.7$ & $1.7\pm0.2$ & $12.31\pm3.47$ & $5.15\pm0.78$ & $7.5\pm15.2$ & C & 4.7 & 1.077 \\
033210.5$-$275940 & 53.043916 & $-$27.994673 & $1.1\pm0.3$ & $0.7\pm0.1$ & $13.00\pm4.05$ & $3.92\pm0.46$ & $178.6\pm10.5$ & C & 4.4 & 1.375 \\
\enddata
\tablecomments{Table \ref{tab:catalog} is published in its entirety in machine-readable format. A subset spanning the flux and S/N range of the catalog is shown here for guidance regarding form and content. Full column definitions, including the island background statistics and per-source R.A./Decl. fitting uncertainties not shown here, are given in Appendix~\ref{sec:catalog_columns}. Maj, Min, and P.A. are fitted (not deconvolved) Gaussian parameters. Compactness values $>1$ indicate a $\sim$97\% probability of being unresolved (Section~\ref{sec:compactness}).}
\end{deluxetable*}

\clearpage
\onecolumngrid

\section{Gaussian Components of Multi-Component Sources}
\label{app:gaussians}

Table~\ref{tab:gaussians} presents the fitted Gaussian components for the 121 sources classified as multi-component ($\texttt{S\_Code}=$ `M') in the main catalog (Table~\ref{tab:catalog}), comprising 363 individual components in total. Table~\ref{tab:gaussians} shows a representative sample; the full table is released in machine-readable format. As in Table~\ref{tab:catalog}, the tabulated coordinates include the astrometric correction described in Section~\ref{sec:astrometry}, and the tabulated flux densities include the primary beam correction described in Section~\ref{sec:pbcor}. Note that the tabulated flux errors are PyBDSF fitting uncertainties only and do not include the overall $\sim$8\% VLITE flux density scale uncertainty for the final deep image (Section~\ref{sec:calib}). Similarly, the tabulated PyBDSF fitting coordinate uncertainties do not include the $1^{\prime\prime}$ systematic positional uncertainty (Section~\ref{sec:astrometry}). The catalog columns are as follows.

\begin{description}

\item[\texttt{Gaus\_id}] Unique identifier for the Gaussian component.

\item[\texttt{Name}] Name of the parent source in the main catalog (Table~\ref{tab:catalog}) to which this component's flux contributes.

\item[\texttt{RA}] Component right ascension, in degrees.

\item[\texttt{e\_RA}] $1\sigma$ fitting error on the right ascension, in arcsec

\item[\texttt{DEC}] Component declination, in degrees.

\item[\texttt{e\_DEC}] $1\sigma$ fitting error on the declination, in arcsec

\item[\texttt{Total\_flux}] Fitted total, integrated Stokes I flux density of the component, in mJy.

\item[\texttt{e\_Total\_flux}] 1$\sigma$ fitting error on the total flux density, in mJy.

\item[\texttt{Peak\_flux}] Fitted peak Stokes I flux density per beam of the component, in mJy~bm$^{-1}$.

\item[\texttt{e\_Peak\_flux}] 1$\sigma$ fitting error on the peak flux density per beam, in mJy~bm$^{-1}$.

\item[\texttt{Maj}] FWHM of the component major axis, in arcsec.

\item[\texttt{e\_Maj}] 1$\sigma$ error on the FWHM of the major axis, in arcsec.

\item[\texttt{Min}] FWHM of the component minor axis, in arcsec.

\item[\texttt{e\_Min}] 1$\sigma$ error on the FWHM of the minor axis, in arcsec.

\item[\texttt{PA}] Position angle of the component major axis measured east of north, in degrees.

\item[\texttt{e\_PA}] 1$\sigma$ error on the position angle of the major axis, in degrees.
\end{description}

\begin{deluxetable*}{lcccccccccc}
\rotate
\tablecaption{Gaussian Components of Multi-Component VLITE HUDF Sources (Sample) \label{tab:gaussians}}
\tablewidth{0pt}
\tablehead{
\colhead{Gaus\_id} & \colhead{Name} & \colhead{R.A.} & \colhead{e\_R.A.} & \colhead{Decl.} & \colhead{e\_Decl.} & \colhead{Total Flux} & \colhead{Peak Flux} & \colhead{Maj} & \colhead{Min} & \colhead{P.A.} \\
\colhead{} & \colhead{(VLITE-A J)} & \colhead{(deg)} & \colhead{(arcsec)} & \colhead{(deg)} & \colhead{(arcsec)} & \colhead{(mJy)} & \colhead{(mJy bm$^{-1}$)} & \colhead{(arcsec)} & \colhead{(arcsec)} & \colhead{(deg)}
}
\startdata
1 & 033711.1$-$283118 & 54.297299 & 0.297 & $-$28.521244 & 0.380 & 12.4 $\pm$ 1.3 & 4.4 $\pm$ 0.3 & 11.22 $\pm$ 0.96 & 8.36 $\pm$ 0.61 & 26.9 $\pm$ 9.9 \\
2 & 033711.1$-$283118 & 54.295977 & 0.047 & $-$28.522362 & 0.154 & 15.7 $\pm$ 0.7 & 11.2 $\pm$ 0.3 & 9.47 $\pm$ 0.36 & 4.95 $\pm$ 0.11 & 178.6 $\pm$ 1.3 \\
3 & 033707.6$-$272736 & 54.282678 & 0.144 & $-$27.460082 & 0.295 & 20.3 $\pm$ 1.1 & 5.7 $\pm$ 0.2 & 14.42 $\pm$ 0.71 & 8.25 $\pm$ 0.31 & 167.9 $\pm$ 2.6 \\
4 & 033707.6$-$272736 & 54.281592 & 0.115 & $-$27.460223 & 0.094 & 18.3 $\pm$ 0.8 & 9.1 $\pm$ 0.3 & 8.80 $\pm$ 0.27 & 7.66 $\pm$ 0.21 & 103.8 $\pm$ 8.3 \\
5 & 033707.6$-$272736 & 54.280524 & 0.022 & $-$27.460085 & 0.056 & 39.3 $\pm$ 0.6 & 23.3 $\pm$ 0.2 & 9.78 $\pm$ 0.13 & 5.78 $\pm$ 0.05 & 1.6 $\pm$ 0.7 \\
6 & 033707.6$-$272736 & 54.284157 & 0.033 & $-$27.460347 & 0.093 & 17.2 $\pm$ 0.5 & 12.8 $\pm$ 0.2 & 8.84 $\pm$ 0.22 & 5.08 $\pm$ 0.08 & 179.4 $\pm$ 1.1 \\
7 & 033703.5$-$271635 & 54.264195 & 0.369 & $-$27.275965 & 0.893 & 1.1 $\pm$ 0.4 & 1.1 $\pm$ 0.2 & 7.80 $\pm$ 2.18 & 4.28 $\pm$ 0.65 & 15.4 $\pm$ 9.8 \\
8 & 033703.5$-$271635 & 54.265459 & 0.549 & $-$27.277052 & 0.989 & 1.1 $\pm$ 0.5 & 0.9 $\pm$ 0.2 & 7.73 $\pm$ 2.43 & 5.07 $\pm$ 1.08 & 160.6 $\pm$ 19.9 \\
\enddata
\tablecomments{Table~\ref{tab:gaussians} is published in its entirety in machine$-$readable format. A subset is shown here for guidance regarding form and content. Full column definitions are given in Appendix~\ref{app:gaussians}. Maj, Min, and P.A. are fitted (not deconvolved) Gaussian parameters.}
\end{deluxetable*}

\end{document}